\documentclass[fleqn,10pt]{wlscirep}
\usepackage[utf8]{inputenc}
\usepackage[T1]{fontenc}

\usepackage[flushleft]{threeparttable}
\usepackage{adjustbox}

\usepackage{color}

\usepackage{aas_macros}

\usepackage{amsmath}
\usepackage{amsfonts}
\usepackage{amssymb}
\usepackage{graphicx}
\usepackage{subcaption}
\usepackage{csquotes}
\usepackage{verbatim}

\usepackage{setspace}
\usepackage{xcolor}
\usepackage{float}

\def\sgr{{Sgr~dSph~}}
\def\fermi{{{\em Fermi}-LAT}}
\def\pairs{{$e^\pm$}}
\def\gr{{$\gamma$-ray}}

\newcommand{\appropto}{\mathrel{\vcenter{
  \offinterlineskip\halign{\hfil$##$\cr
    \propto\cr\noalign{\kern2pt}\sim\cr\noalign{\kern-2pt}}}}}

\newcommand{\msun}{M_{\odot}}

\newcommand{\aref}[1]{\hyperref[#1]{Appendix~\ref{#1}}}

\usepackage[backend=biber, style=nature, eprint=false, doi=false]{biblatex}

\usepackage{jabbrv}

\AtEveryBibitem{
  \clearfield{month}
}

\DeclareFieldFormat{journal}{ \itshape #1 }
\DefineBibliographyStrings{english}{andothers = {\normalfont et\addabbrvspace al\adddot}}

\makeatletter
\AtEveryCitekey{%
  \ifcsundef{blx@entry@refsegment@\the\c@refsection @\thefield{entrykey}}
    {\csnumgdef{blx@entry@refsegment@\the\c@refsection @\thefield{entrykey}}{\the\c@refsegment}}
    {}}
    
\defbibcheck{onlynew}{%
  \ifnumless{0\csuse{blx@entry@refsegment@\the\c@refsection @\thefield{entrykey}}}{\the\c@refsegment}
    {\skipentry}
    {}}
\makeatother

\title{Gamma-ray emission from the Sagittarius Dwarf Spheroidal galaxy due to millisecond pulsars}

\author[1,10,*,+]{Roland M.~Crocker}
\author[2,3,$\dagger$,+]{Oscar Macias}
\author[1]{Dougal Mackey}
\author[1]{Mark R.~Krumholz}
\author[2,3]{Shin’ichiro Ando}
\author[4,3]{Shunsaku Horiuchi}
\author[5]{Matthew G.~Baring}
\author[6]{Chris Gordon}
\author[7]{Thomas Venville}
\author[7]{Alan R.~Duffy}
\author[8,9,10]{Rui-Zhi Yang}
\author[11,12]{Felix Aharonian}
\author[11]{J.~A. Hinton}
\author[4]{Deheng Song}
\author[13]{Ashley J.~Ruiter}
\author[14]{Miroslav D. Filipovi{\'c}}

\affil[1]{Research School of Astronomy and Astrophysics, Australian National University, Canberra 2611, A.C.T., Australia}
\affil[2]{GRAPPA $-$ Gravitational and Astroparticle Physics Amsterdam, University of Amsterdam, Science Park 904, 1098 XH Amsterdam, The Netherlands}
\affil[3]{Kavli IPMU (WPI), UTIAS, The University of Tokyo, Kashiwa, Chiba 277-8583, Japan}
\affil[4]{Center for Neutrino Physics, Department of Physics, Virginia Tech, Blacksburg, Virginia 24061, USA}
\affil[5]{Department of Physics and Astronomy - MS 108, Rice University, 6100 Main Street, Houston, Texas 77251-1892, USA}
\affil[6]{School of Physical and Chemical Sciences, University of Canterbury, Christchurch, New Zealand}
\affil[7]{Centre for Astrophysics and Supercomputing, Swinburne University of Technology,
PO Box 218, Hawthorn, VIC 3122, Australia}
\affil[8]{Department of Astronomy, School of Physical Sciences, University of Science and Technology of China, Hefei, Anhui 230026, China}
\affil[9]{CAS Key Labrotory for Research in Galaxies and Cosmology, University of Science and Technology of China, Hefei, Anhui 230026, China}
\affil[10]{School of Astronomy and Space Science, University of Science and Technology of China, Hefei, Anhui 230026, China}
\affil[11]{Max-Planck-Institut für Kernphysik, Saupfercheckweg 1, 69117 Heidelberg, Germany}
\affil[12]{Dublin Institute for Advanced Studies, 31 Fitzwilliam Place, Dublin 2, Ireland}
\affil[13]{School of Science, University of New South Wales Canberra, The Australian Defence Force Academy, 2600 ACT, Canberra, Australia}
\affil[14]{Western Sydney University, Locked Bag 1797, Penrith South DC, NSW 2751, Australia}

\affil[*]{rcrocker@fastmail.fm}
\affil[$\dagger$]{o.a.maciasramirez@uva.nl}
\affil[+]{These authors contributed equally to this work.}

\newrefsegment

\begin{abstract}
\textbf{The Fermi Bubbles are giant, $\gamma$-ray emitting lobes emanating from the nucleus of the Milky Way\cite{Su2010,Ackermann2014}
discovered in $\sim$ 1-100 GeV data collected by the Large Area Telescope on board
the  {\em Fermi} Gamma-Ray Space Telescope\cite{Atwood2009}.
Previous work\cite{Su2012} has revealed substructure within the Fermi Bubbles  that has been interpreted as a signature of collimated outflows from the Galaxy’s super-massive black hole. 
Here we show 
via a spatial template analysis
that 
much of the $\gamma$-ray emission associated to
the brightest region of substructure  
-- the so-called {\em cocoon} -- 
is 
likely
due to the Sagittarius dwarf spheroidal (Sgr dSph) galaxy.
This large Milky Way satellite is viewed through the Fermi Bubbles from the position of the Solar System. 
As a tidally and ram-pressure stripped remnant, the Sgr dSph has no on-going star formation, but we 
nevertheless
demonstrate that 
the dwarf’s millisecond pulsar (MSP) population can plausibly supply the
\gr \ signal that our analysis associates to its stellar template.
The measured spectrum is
naturally explained by inverse Compton scattering of cosmic microwave background photons by high-energy electron-positron  pairs injected by MSPs belonging to the Sgr dSph, combined with these objects' magnetospheric emission.
This finding plausibly suggests that MSPs produce significant \gr \ emission amongst old stellar populations, potentially confounding indirect dark matter searches in regions such as the Galactic Centre, the Andromeda galaxy, and other massive Milky Way dwarf spheroidals.
}
\end{abstract}
\begin{document}

\flushbottom
\maketitle
 
\thispagestyle{empty}

\begin{figure}[H]
 \includegraphics[width=1.\linewidth]{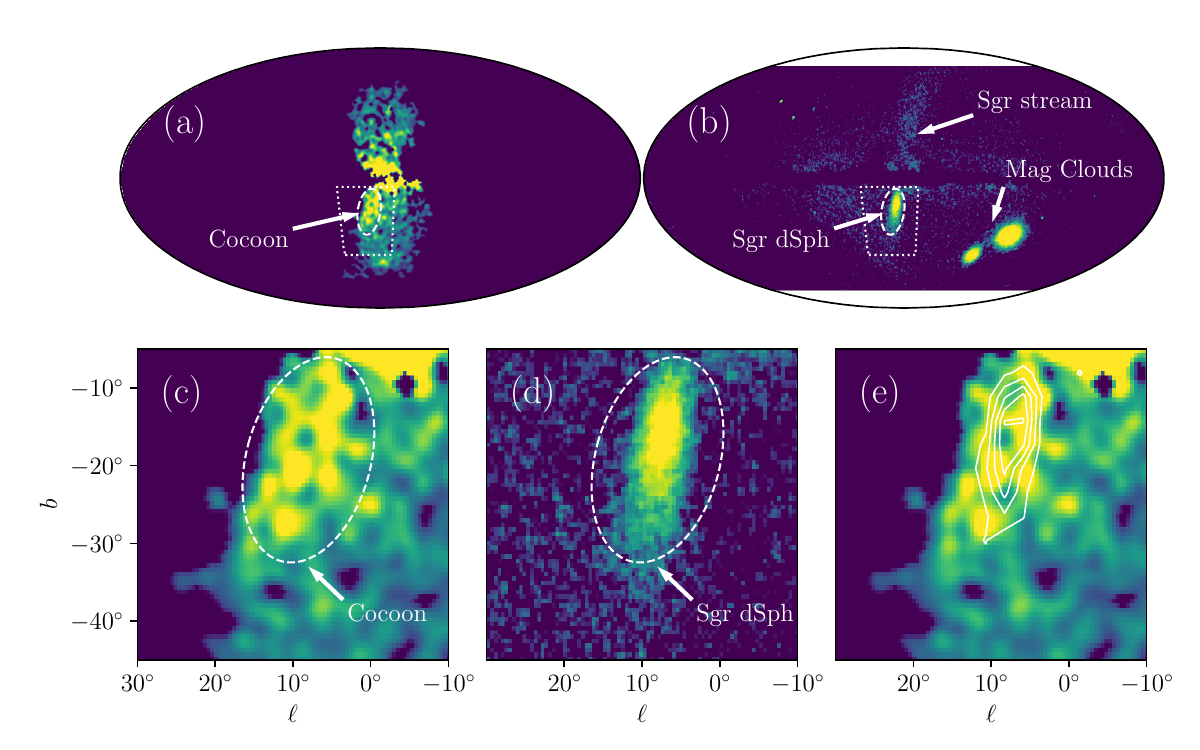}
    \caption{
    {\bf The Fermi Bubbles, including the Cocoon sub-structure, and the Sgr dSph galaxy.}
    \textbf{
    Panels (a) and (c)} display 
    the $\gamma$-ray spatial template for the Fermi Bubbles\cite{Ackermann2014} in arbitrary units with linear colour scale, highlighting the cocoon.
    \textbf{Panels (b) and (d)}
    show the angular density of RR Lyrae stars with line-of-sight distances $>20$ kpc from the {\it Gaia} Data Release 2 (DR2), in arbitrary units with logarithmic scaling; the Sgr dSph, Sgr stream, and the Large and Small Magellanic Clouds are clearly visible. The proper motion of the \sgr is upwards in this figure. The dashed ellipses in panels (a)-(d) mark the same coordinates in each panel, and highlight both the cocoon and the Sgr dSph.
    \textbf{Panel (e)} shows contours of RR Lyrae surface density overlaid on the Fermi Bubbles template shown as the coloured background.
    Panels \textbf{(a) and (b)} are all-sky views
    in Mollweide projection in
    Galactic coordinates of longitude $\ell$ and latitude $b$ with east to the left, with the ROI marked by the dotted box.
    Panels \textbf{(c)--(e)} are in a cylindrical projection and zoom in on the region of interest.
    }
    \label{fig:SgrdSphOverlay}
\end{figure}

\noindent
Early analysis of data from the \textit{Fermi} Large Area Telescope (\textit{Fermi}-LAT)  identified two
counter-propagating, co-linear \gr \ substructures within the Fermi Bubbles (FBs; \autoref{fig:SgrdSphOverlay}a), a {\it jet} in the northern Galactic hemisphere and {\it cocoon}  in the south\cite{Su2012}; 
subsequent, independent analyses\cite{Ackermann2014,Selig2015}  have only confirmed the existence of the latter.
Since the cocoon is contained within the solid angle of the surrounding FBs and exhibits a similar $\gamma$-ray spectrum,
it is natural to propose they share a common origin.
However, the cocoon is also spatially coincident with the core of the Sagittarius dwarf spheroidal galaxy (Sgr dSph \cite{Ibata1994}; \autoref{fig:SgrdSphOverlay}b), a satellite of the Milky Way that is in the process of being accreted and destroyed, as tidal forces gradually strip stars out of its core into elongated streams\cite{Belokurov2006}. 
The chance probability of such an  alignment is low, $\sim 1$\% (see the Supplementary Information; S.I.~sec.~1), even before accounting for the fact that the cocoon and the \sgr have similar shapes and
orientations, and that the Sgr dSph is both one of the nearest and most massive ($d=26.5$ kpc, $M\sim 10^8$ M$_\odot$; \cite{Vasiliev2020, Vasiliev2021}) Milky Way satellites and
has the largest mass divided by distance squared of \textit{any} astronomical object not yet detected in $\gamma$-rays.

We therefore consider emission from the \sgr as an alternative origin for the cocoon. In order to test this possibility, we fit the \gr~emission observed by \textit{Fermi}-LAT over a region of interest (ROI) containing the cocoon 
via template analysis.
In our {\it baseline} model these templates include only known point sources and sources of Galactic diffuse \gr~emission.
We contrast the baseline with a {\it baseline + Sgr dSph}
model that invokes these same templates plus an additional template constructed to be spatially coincident with the bright stars of the Sgr dSph (Extended Data (E.D.) \autoref{fig:Stellartemplates} and S.I.~\autoref{fig:profile_longaxis}); full details of the fitting procedure are provided in Methods and S.I.~sec.~3. 
Using the best motivated 
choice of templates, we find that the baseline + Sgr dSph model is preferred at 
$8.1\sigma$ significance over the baseline model. We also repeat the analysis for a wide range of alternative templates for both Galactic diffuse emission and for the Sgr dSph (\autoref{tab:loglikelihood}) and obtain $>5\sigma$ detections for all combinations but one.
Moreover, even this is an extremely conservative estimate, because our baseline model uses a structured template for the FBs that absorbs some of the signal that is spatially coincident with the Sgr dSph into a structure of unknown origin. If we follow the method recommended by the Fermi collaboration \cite{Ackermann2014} and use a flat FB template in our analysis, the significance of our detection of the Sgr dSph is always $>14\sigma$. Despite this, for the remainder of our analysis we follow the most conservative choice by using the  structured template in our baseline model. In Methods, we also show that our analysis passes a series of validation tests: the residuals between our best-fitting model and the data are consistent with photon counting statistics (E.D.~\autoref{fig:fitvalidation} and \autoref{fig:Residuals}), our pipeline reliably recovers synthetic signals superimposed on a realistic background (E.D.~\autoref{fig:injectionrecovery}), fits using a template tracing the stars of the Sgr dSph yield significantly better results than fits using purely geometric templates (S.I.~\autoref{tab:geometric_templates}), and if we artificially rotate the Sgr dSph template on the sky, the best-fitting position angle is very close to the actual one (E.D.~\autoref{fig:rotationAndTranslationTests}).
By contrast, if we displace the Sgr dSph template,
we find moderate ($4.5\sigma$ significance) evidence that the best-fitting position 
is $\sim 4^\circ$ from the true position,
in a direction very closely aligned with the dwarf galaxy's direction of travel (E.D. \autoref{fig:rotationAndTranslationTests}); this plausibly represents a small, but real and expected (as explained below) physical offset between the stars and the \gr~emission.

\begin{figure}[H]
    \centering
        \includegraphics[width=\linewidth]{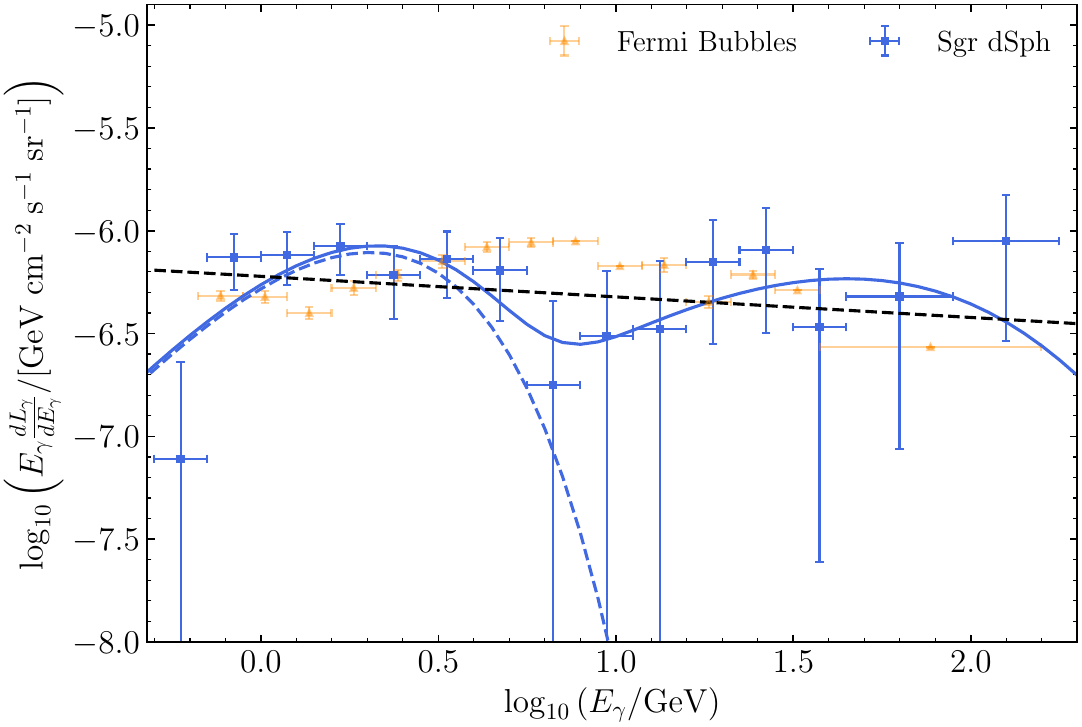}
\caption{ 
{\bf Measured $\gamma$-ray spectral brightness distributions of the signal associated to} the \sgr template and the surrounding Fermi Bubbles. 
The black, dashed line 
shows a differential number flux obeying
$dN_\gamma/dE_\gamma \propto E_\gamma^{-2.1}$.
These data 
are as obtained by us in our \fermi \ data analysis as described in Methods. 
We have converted luminosities to surface brightnesses 
adopting source solid angles of $\Omega_{\rm Sgr \ dSph} = 9.6 \times 10^{-3}$ sr, 
and $\Omega_{\rm FB} = 0.49$ sr, with the latter set by the $40^\circ \times 40^\circ$ region of interest (ROI), not the intrinsic sizes of the Bubbles (which are larger than the ROI).
Error bars show $1\sigma$ errors; for the Sgr dSph, the error bars incorporate both statistical and systematic errors added in quadrature.
The smooth blue curves show (solid) the best fit combined (magnetospheric + IC) and (dashed) the best fit magnetospheric spectra.
}
\label{fig:luminosities}
\end{figure}

The directly-measured flux from the Sgr dSph, derived from our fiducial choice of templates, corresponds to a luminosity of $(3.8 \pm 0.6) \times 10^{36}$ erg/s (1$\sigma$ error) 
for \gr \ photons in the range from 0.5 to 150 GeV 
(equivalently $\sim 4 \times 10^{28}$ erg/s/$\msun$). Over this range the spectrum is approximately described by a hard power law $dF_\gamma/dE_\gamma \appropto E_\gamma^{-2.1}$ (\autoref{fig:luminosities}).
There is no evidence for a cut-off at high energies.
We show in E.D.~\autoref{fig:SgrSpecVar} that this spectral shape is qualitatively insensitive to the choice of foreground templates, and E.D.~\autoref{fig:totalspectra} demonstrates that the spectra we recover for the various foregrounds within the ROI
remain physically plausible when we introduce a \sgr template.

Since our template fits 
plausibly
suggest that there is a real \gr~emission component tracing the Sgr dSph, a natural next question is what mechanism could be responsible for producing it. 
The core of the Sgr dSph is the remnant of a once much more massive galaxy.
Tidal and ram pressure stripping removed its gas and caused it to cease forming stars $2-3$ Gyr ago \cite{Siegel2007}, though
it did experience punctuated bursts of star formation \cite{Weisz2014} -- triggered by its crossings through the Galactic plane \cite{Tepper-Garcia2018} -- up to 
that time.
In the MW, the dominant source of diffuse \gr~emission is collisions between 
(hadronic) cosmic rays (CRs)   and ambient 
instellar medium
(ISM) gas nuclei\cite{Strong2010}, but this mechanism cannot operate in the Sgr dSph, which lacks both `target' gas with which CRs could interact, and supernova explosions from young, massive stars to accelerate hadronic CRs in the first place.
%
Stellar \gr \ emission is also ruled out: 
while our Sun is a source of $\sim 100$ GeV $\gamma$-rays, this emission is again dominantly due to collisions between  hadronic CRs from the wider Galaxy and Solar gas;
\gr \ emission from non-thermal particles accelerated by the Sun itself only extends to 4 GeV \cite{Linden2018}.
This leaves two possibilities for the   $\gamma$-ray signal our template analysis associates to the Sgr dSph:
it is created
from the self-annihilation of 
dark matter (DM) particles in the dwarf's DM halo, or
by millisecond pulsars (MSPs) deriving from the stars of the Sgr dSph. The former is unlikely because the $\gamma$-ray
signal largely traces the stars of the dwarf, while
N-body simulations \cite{Tepper-Garcia2018}
show that the Milky Way's tidal field 
will have
overwhelmingly dispersed the progenitor galaxy's original DM halo into the stream over its orbital history.

MSPs, by contrast, should follow the same spatial distribution as the rest of the stellar population, have a  spin-down timescale $\gtrsim \mathcal{O}$(Gyr), long enough to be compatible with the most recent episodes of Sgr dSph star formation, and
radiate some part of their magnetic dipole luminosity into
$\gamma$-rays.
However, there are two significant challenges to this scenario: 
First, the inferred $\gamma$-ray luminosity per unit stellar mass is much larger ($\gtrsim 10 \times $)  for the \sgr than for some
other systems whose detected \gr \ emission is plausibly dominated by MSPs including the Galactic Bulge \cite{Abazajian2011,Macias2018,Bartels2018,Macias2019,Gautam2021} and
Andromeda\cite{Ackermann2017,Eckner2018} (M31), the giant spiral galaxy  nearest to the Milky Way (although it is smaller than that observed for globular clusters -- see \autoref{fig:LgammaOvrMstar}).
Second, the hard, $\appropto E_\gamma^{-2.1}$
spectrum of the \sgr (\autoref{fig:luminosities}) does not resemble the classic $\sim$ few GeV bump  (in the spectral energy distribution) of the magnetospheric \gr \ signal detected from 
individual MSPs 
or the globular clusters (GCs) that host populations of MSPs: e.g. \cite{Song2021}.

However, both of these challenges can be overcome by considering how the stellar population and interstellar environment of the Sgr dSph differ from other systems. With regard to stars, those in the Sgr dSph are both younger and more metal-poor than those of M31 or the Galactic Bulge; metal-poor stellar systems are expected to produce more MSPs per stellar mass \cite{Ruiter2019}, and $\sim 7-8$ Gyr-old MSPs (the rough age of the Sgr dSph population) are expected to be significantly brighter than $10-12$ Gyr-old ones (the ages of stellar populations in the Bulge and the core of M31) \cite{Gautam2021}. In S.I.~sec.~5 we 
show that the best-fit value for the $\gamma$-ray luminosity of the Sgr dSph is fully consistent with both theoretical predictions and with observations of other $\gamma$-ray emitting old stellar populations once age and metallicity are taken into account. On the basis of stellar population synthesis models, we estimate that the $\gamma$-ray luminosity of the Sgr dSph is produced by $\sim 650$ MSPs.

With regard to environment, note that while the spectrum of the \sgr does not resemble an MSP magnetospheric signal, it does resemble inverse Compton (IC) emission from the up-scattering (by a CR electron-positron  population; \pairs) of ambient light  which, for the Sgr dSph, is dominated by the Cosmic Microwave Background (CMB). We also know that MSPs produce \pairs~with energies of at least a few TeV, since these are the particles that ultimately drive the observed GeV MSP $\gamma$-ray photospheric emission. Some of these \pairs~will give up all their energy within the MSP magnetosphere. However, given the expected absence of wind nebulae or supernova remnants surrounding these old, low luminosity objects \cite{Sudoh2020}, many will freely escape both magnetosphere and MSP environs into the larger \sgr~environment \cite{Baring2011,Venter2015} where they can IC up-scatter CMB photons.
In an environment like Andromeda or the Galactic Bulge, this IC signal will be weak (albeit detectable in the case of the Bulge according to \cite{Gautam2021}), because much of the escaping \pairs~energy will be lost to synchrotron rather than IC radiation. In an ultra-gas poor system like the Sgr dSph, however, we expect the ISM magnetic field to be far weaker than in a gas-rich galaxy (\cite{Regis2015}; also see Methods) with an energy density significantly smaller than that in the CMB; thus radiative losses from MSP-escaping \pairs \ are overwhelmingly into hard-spectrum IC $\gamma$-rays rather than 
(radio to X-ray) 
synchrotron radiation. 
Consistent with this explanation, globular clusters -- which are also gas-poor and weakly-magnetized -- represent another environment where MSP-driven $\gamma$-ray emission seems to sometimes include a significant IC component\cite{Song2021}. We formalize this intuitive argument in Methods, where we show that the spectrum of the \sgr is extremely well fit as a combination of IC and magnetospheric radiation with self-consistently related spectral parameters. This scenario also explains why the $\gamma$-ray signal is
displaced
$\sim 4^\circ$,  or about 1.9 kpc (E.D.~\autoref{fig:rotationAndTranslationTests}, right),
from the center of the Sgr dSph; the dwarf's
Northward proper motion
\cite{delPino2021} means this displacement is backwards along its path. As the \sgr~plunges through the Milky Way halo, the magnetic field around it will be elongated into a magnetotail oriented backwards along its trajectory, and \pairs~emitted into the dwarf will be trapped by these magnetic field lines, leading them to accumulate and emit in a position that trails the Sgr dSph, exactly as we observe. We offer a more quantitative evaluation of this scenario in  S.I.~sec.~4.

There are some caveats to our results that the reader should note.
First, in common with 
other {\it Fermi}-LAT data analyses of diffuse emission from extended regions,
it is evident that our model, while it is very good, does not reproduce the data
accurate down to the level of Poisson noise over the entire ROI (see discussion in S.I.~sec.~3). 
Indeed, E.D.~Fig.~3 shows that there are structured residuals within the ROI, though we note that the strongest of these are at the edges of the ROI and not coincident with the Sgr dSph.
We do not believe, therefore, that these residuals indicate that the detection of the signal connected to the \sgr stellar template made in our \gr \ analysis
is spurious, nor that the spectrum we measure is likely to be in significant error (see E.D.~Fig.~4).
Rather, we suspect that the structured residuals point to the existence of still mis-modelled sub-structure in the Fermi Bubbles that is completely unrelated to the Sgr dSph.
Thus, while we argue on the basis of our analysis that much of the cocoon substructure is likely emission from the Sgr dSph, we do not claim to explain {\it all} Fermi Bubbles substructure.

This point connects to a second caveat: 
We are aware of no independent, multi-wavelength (non \gr ) evidence for the existence of a well-defined, nuclear jet or jets on angular scales comparable to the Fermi Bubbles.
Thus,
in distinction to the case presented by the \sgr (where we can construct from independent, multi-wavelength data a spatial template to incorporate into our \gr \ analysis), 
we cannot construct any definitive, {\it a priori} jet template.
While we argue that this is actually a weakness of the jet hypothesis, it nevertheless is true that we cannot cannot via a formal statistical analysis rule out the presence of \gr \ sub-structure in the Fermi Bubbles that is connected to a nuclear jet.

Taking note of all the above, there are a number of potential implications of
the discovery of a \gr \ signal associated to the \sgr 
stellar template to follow up.
Firstly, our results  motivate the introduction of stellar templates into the analysis of data from all \gr \ resolved galaxies (M31, Large and Small Magellanic Clouds) to probe the contribution of MSPs. 
Such studies may confirm (or not) whether the rather strong signal our analysis associates to the \sgr stellar template can indeed be explained
reasonably via MSP emission (see E.D.~Fig.~9 and S.I. section 5).
Secondly, our study lends support to the argument \cite{Sudoh2020} that MSPs contribute significantly to the energy budget of CR \pairs \ in galaxies with low specific star-formation rates. 
Third, we show in the SI that a 
direct extrapolation of the Sgr dSph MSP $\gamma$-ray luminosity per unit mass to other nearby dSph galaxies suggests that they could have considerably larger astrophysical $\gamma$-ray signatures than previous estimates; we report our revised estimates in S.I.~\autoref{tab:dsph_fluxes} for the sample of ref.~\cite{Winter2016}. 
These signals are large enough that some are potentially detectable via careful analysis of Pass8 (15-year) \textit{Fermi}-LAT data. 
Conversely, these brighter astrophysical signatures represent a larger-than-expected background with which searches for DM annihilation signals 
(due to putative WIMPs in the tens of GeV mass range)
must contend, and potentially swamp DM signals in some nearby dwarfs. 
We emphasise that these are not predictions {\it per se} but, rather, naive extrapolations that do not account for peculiarities of the \sgr with respect to other dSphs that may render it peculiarly \gr \ efficient (e.g., its relatively recent star formation).
These extrapolations do, nevertheless, motivate further work to pin down in detail how the \gr \ luminosity of an MSP population scales with gross parameters of the host stars (mass, age, metallicity, etc).


Supplementary Information is available for this paper. Correspondence and requests for materials should be addressed to RMC or OM. 

\clearpage


\section*{Methods}
\label{sec:Methods}

Our analysis pipeline consists of three steps: (1) data and template selection, (2) fitting, and (3) spectral modeling. 

\subsection*{
Data and template selection
}
\label{sec:fermidata}

We use eight years of LAT data,  selecting \texttt{Pass 8 UltraCleanVeto} class events in the energy range from 500 MeV to 177.4 GeV. We choose the limit at low energy to mitigate both the impact of \gr \ leakage from the Earth's limb  and the increasing width of the point-spread function at lower energies.
We spatially bin the data to a resolution of $0.2^\circ$, and divide it into 15 energy bins; the 13 lowest-energy of these are equally spaced in log energy, while the 2 highest-energy are twice that width in order to improve the signal to noise.
We select data obtained over the same observation period as that used in the construction of the Fourth Fermi Catalogue (4FGL)\cite{Fermi-LAT:4FGL} (August 4, 2008 to August 2, 2016).
The region of interest (ROI) of our analysis is a square region defined by $-45^\circ\leq b \leq -5^\circ$, and $30^\circ \geq \ell \geq -10^\circ$ (\autoref{fig:SgrdSphOverlay}). 
This sky region fully contains the Fermi cocoon substructure but avoids the Galactic plane ($|b|\leq 5^\circ$) where uncertainties are largest. Because the ROI is of modest size, we allow the Galactic diffuse emission (GDE)
templates greater freedom to reproduce potential features in the data. 
We carry out all data reduction and analysis using the standard  
\textsc{Fermitools v1.0.1}
software package (available from 
\url{https://github.com/fermi-lat/Fermitools-conda/wiki}).
We model the performance of the LAT with the \texttt{P8R3\_ULTRACLEANVETO\_V2} Instrument Response Functions (IRFs).

We fit the spatial distribution of the ROI data as the sum of a series of templates for different components of the emission. For all the templates we consider, we define a ``baseline'' model that includes only known point and diffuse emission sources, to which we compare a ``baseline + Sgr dSph'' model that includes those templates plus the Sgr dSph. Our baseline models, following the approach of Ref.~\cite{Abazajian:2020}, contain the following templates: (1) diffuse isotropic emission, (2) point sources, (3) emission from the Sun and Moon, (4) Loop I, (5) the Galactic Centre Excess, (6) Galactic cosmic ray-driven hadronic and bremsstrahlung emission, (7) inverse Compton emission, and (8) the \textit{Fermi} Bubbles; baseline + Sgr dSph models also include a Sgr dSph template.

Our templates for the first five emission sources are straightforward, and we adopt a single template for each of them throughout our analysis. Since our data selection is identical to that used to construct the 4FGL, we adopt the standard isotropic background and point source models provided as part of the catalogue \cite{Fermi-LAT:4FGL}, \texttt{iso$_{-}$P8R3$_{-}$ULTRACLEANVETO$_{-}$V2$_{-}$v1.txt}, and \texttt{gll\_psc\_v20.fit}, respectively; the latter includes 177 $\gamma$ ray point sources within our ROI. We similarly adopt the standard Sun and Moon templates provided. For the foreground structure Loop I, we adopt the model of Ref.~\cite{Wolleben:2007}. Finally, given that the low-latitude boundary of our ROI overlaps with the spatial tail of the Galactic Centre Excess (GCE), we include the `Boxy Bulge' template of Ref.~\cite{Freudenreich1998}, which has been shown \cite{Macias2018,Bartels2018,Macias2019} to provide a good description of the observed GCE away from the nuclear Bulge region (which is outside our ROI). The inclusion of this template in our ROI model has only a small impact on our results.

The remaining templates require more care. The dominant source of $\gamma$-rays within the ROI is hadronic and bremsstrahlung emission resulting from the interaction of Milky Way cosmic ray (CR) protons and electrons with interstellar gas; the emission rate is proportional to the product of the gas density and the CR flux. We model this distribution using three alternative approaches. Our preferred approach follows that described in Ref.~\cite{Macias2018}. We assume that the spatial distribution of $\gamma$-ray emission traces the gas distribution from the hydrodynamical model of Ref.~\cite{Pohl2008}, which gives a more realistic description of the inner Galaxy than alternatives. To normalise the emission, we divide the Galaxy into four rings spanning the radial ranges $0-3.5$ kpc, $3.5-8.0$ kpc, $8.0-10.0$ kpc, and $10.0-50.0$ kpc, within which we treat the emission per unit gas mass in each of our 15 energy bins as a constant to be fit. We refer to the template produced in this way as the ``HD'' model. Our first alternative is to use the same procedure of dividing the Galaxy into rings, but describe the gas distribution within those rings using a template constructed from interpolated maps of Galactic H~\textsc{i} and H$_2$, following the approach described in Appendix B of Ref.~\cite{Ackermann2012}; we refer to this as the ``Interpolated'' approach. Our third alternative, the ``GALPROP'' model, is the SA50 model described by Ref.~\cite{Johannesson:2018bit}, which prescribes the full-sky hadronic CR emission distribution.

We similarly need a model for diffuse, Galactic IC emission -- the second largest source of background -- which is a product of the CR electron flux and the interstellar radiation field (ISRF). As with hadronic emission, we consider four alternative distributions. Our default choice is the SA50 model described by Ref.~\cite{Johannesson:2018bit}, which includes 3D models for the ISRF~\cite{Porter:2017vaa}. We therefore refer to this as the ``3D'' model. However, unlike in Ref.~\cite{Johannesson:2018bit}, we use this model only to obtain the spatial distribution of the emission, not its normalisation or energy dependence. Instead, we obtain these in the same way as for our baseline hadronic emission model, i.e., we divide the Galaxy into four rings and leave the total amount of emission in each ring at each energy as a free parameter to be fit to the data; this approach reduces the sensitivity of our results to uncertainties in the electron injection spectrum and ISRF normalisation. Our three alternatives to this are models ``2D A'', ``2D B'', and ``2D C'', corresponding to models A, B, and C as described by Ref.~\cite{Ackermann:2014usa}, which model IC emission over the full sky under a variety of assumptions about CR injection and propagation, but rely on a 2D model for the ISRF.

The final component of our baseline template is a model for the Fermi Bubbles themselves, which are one of the strongest sources of foreground emission in high latitude regions of the ROI. The FBs are themselves defined as highly statistically-significant and spatially-coherent residuals in the inner Galaxy that remain once other sources are modelled out in all-sky \gr \ analyses. The FBs are not reliably traced by emission at any other wavelength, so we do not have an {\it a priori} model with which to guide the construction of a spatial template of these structures. However, one characteristic that renders the FBs distinct from other large angular scale diffuse $\gamma$-ray structures is their hard \gr \ spectrum. Indeed, the state-of-the-art, {\it structured} spatial template for them generated by the {\it Fermi} Collaboration\cite{Ackermann2014} -- the templates one would normally employ in large ROI, inner Galaxy \fermi \ analyses --  were constructed using a spectral component analysis. That study recovered a number of regions of apparent substructure within the solid angle of the FBs,  most notably substructure overlapping the previously-discovered\cite{Su2012,Selig2015} ``cocoon'' which, as we have discussed here, is largely coincident with the Sgr dSph. Of course, a potential issue with constructing a phenomenological, spectrally-defined model for the FBs is that, if there happens to be an extended, spectrally-similar source coincident with the FBs, it will tend to be incorporated into the template. For this reason Ref.~\cite{Ackermann2014} suggest using a flat FB template when searching for new structures. Despite this proposal, our default analysis uses the more conservative choice of a structured FB template. However, we also run tests using an unstructured template for comparison, and to understand the systematic uncertainties associated with the choice of template. We refer to these two cases as the ``U'' (Unstructured) and ``S'' (Structured) FB templates, respectively.

Finally, our baseline + Sgr dSph models require a template for the Sgr dSph. Our templates trace the distribution of bright stars in the dwarf, which we construct from five alternative stellar catalogues, all based on different selections from \textit{Gaia} Data Release 2; we refer to the resulting templates as models I - V, and show them in E.D.~\autoref{fig:Stellartemplates}. Full details on how we construct each of these templates are provided in  S.I.~sec.~2. Model I, our default choice, comes from the catalogue of $2.26\times 10^5$ Sgr dSph candidate member stars from Ref.~\cite{Vasiliev2020}; the majority of the catalogue consists of red clump stars. Model II uses the catalogue of RR Lyrae stars in the Sagittarius Stream from Ref.~\cite{Ibataetal:2020}, which we have down-selected to a sample of 2369 stars whose kinematics are consistent with being members of the Sgr dSph itself. Model III uses the catalogue of $1.31\times 10^4$ RR Lyrae stars belonging to the Sgr dSph provided by Ref.~\cite{Iorio2019}. Finally, models IV and V come from the nGC3 and Strip catalogues of RR Lyrae stars from Ref.~\cite{Ramosetal:2020}; the former contains 675 stars with higher purity but lower completeness, while the latter contains 4812 stars of higher completeness but lower purity.

\subsection*{Fitting procedure}

Our fitting method follows that introduced in Refs.~\cite{Macias2018, Macias2019}, and treats each of the 15 energy bins as independent, thereby removing the need to assume any particular spectral shape for each component and allowing the spectra to be determined solely by the data. Our data to be fit consist of the observed $\gamma$-ray photon counts in each spatial pixel $i$ and energy bin $n$, which we denote $\Phi_{n,i,\rm obs}$, where 
$n$ goes from 1 to 15, and the index $i$ runs over the positions $(\ell_i,b_i)$ of all spatial pixels within the ROI. For a given choice of template, we write the corresponding model-predicted $\gamma$-ray counts as $\Phi_{n,i,\rm mod} = \sum_c \mathcal{N}_{n,c} R_{n,i} \Phi_{c,i}$, where $R_{n,i}$ is the instrument response for each pixel and energy bin (computed assuming an $E^{-2}$ spectrum within the bin), and $\Phi_{c,i}$ is the value of template component $c$ evaluated at pixel $i$; for baseline models, we have a total of 8 components, while for baseline + Sgr dSph models we have 9. Note that $\Phi_{c,i}$ is a function of $i$ but not of $n$, i.e., we assume that the spatial distribution of each template component is the same at all energies, except for the IC templates, for which an energy-dependent morphology is predicted by our GALPROP simulations. Without loss of generality we further normalise each template component as $\sum_i \Phi_{c,i} = 1$, in which case $\mathcal{N}_{n,c}$ is simply the total number of photons contributed by component $c$ in energy bin $n$, integrated over the full ROI; the values of $\mathcal{N}_{n,c}$ are the parameters to be fit. We find the best fit by maximising the usual Poisson likelihood function
\begin{equation}
    \ln\mathcal{L}_n = \sum_{i} \frac{\Phi_{n,i,\rm mod}^{\Phi_{n,i,\rm obs}} e^{-\Phi_{n,i,\rm mod}}}{\Phi_{n,i,\rm obs}!},
    \label{eq:log-likelihood}
\end{equation}
using the \texttt{pylikelihood} routine, the standard maximum-likelihood method in \texttt{FermiTools}. Note that, since each energy bin $n$ is independent, we carry out the likelihood maximisation bin-by-bin.

We perform all fits in pairs, one for a baseline model containing only known emission sources, and one for a baseline + Sgr dSph model containing the same known sources plus a component tracing the Sgr dSph. The set of paired fits we perform in this manner is shown in \autoref{tab:loglikelihood}. We compare the quality of these baseline and baseline + Sgr dSph fits by defining the test statistic $\mathrm{TS}_n = -2\ln(\mathcal{L}_{n,\rm base}/\mathcal{L}_{n,\rm base+Sgr})$; the total test statistic for all energy bins is simply $\mathrm{TS} = \sum_n \mathrm{TS}_n$. We can assign a $p$-value to a particular value of the TS by noting that baseline + Sgr dSph models have 15 additional degrees of freedom compared to baseline models: the value of $\mathcal{N}_{n,c}$ for the component $c$ corresponding to the Sgr dSph, evaluated at each of the  
15 energy bins. In this case, the mixture distribution formula gives\cite{Macias2018}
\begin{equation}
    p(\mathrm{TS}) = 2^{-N} \left[\delta(\mathrm{TS}) + \sum_{n=1}^N \binom{N}{n} \chi^2_n(\mathrm{TS})\right],
\end{equation}
where $N = 15$  is the difference in number of degrees of freedom, 
$\binom{N}{n}$ is the binomial coefficient, $\delta$ is the Dirac delta function, and $\chi^2_n$ is the usual $\chi^2$ distribution with $n$ degrees of freedom. The corresponding statistical significance (in $\sigma$ units) is\cite{Macias2018}:
\begin{equation}
\label{eq:numberofsigmas}
\mbox{Number of $\sigma$}\equiv \sqrt{\rm InverseCDF\left(\chi_1^2,{\rm CDF}\left[p(\mbox{TS}),\hat{{\rm TS}}\right]\right)},
\end{equation}
where (InverseCDF) CDF is the (inverse) cumulative distribution function and the first argument of each of these functions is the distribution function, the second is the value at which the CDF is evaluated, and the total TS is denoted by $\hat{\rm TS}$. For 15 extra degrees of freedom, a 5$\sigma$ detection corresponds to $\mbox{TS}=46.1$. (Additional details of these formulae are given in S.I. Sec.~2 of Ref.~\cite{Macias2018}.) We report values of $\mathcal{L}_{\rm base}$, $\mathcal{L}_{\rm base+Sgr}$, $\mathrm{TS}$, and the significance level for all the templates we try in \autoref{tab:loglikelihood}.

A final step in our fitting chain is to assess the uncertainties. For our default choice of baseline + Sgr dSph model (first row in \autoref{tab:loglikelihood}), our maximum likelihood analysis returns the central value $\mathcal{N}^{\rm def}_n$ on the total $\gamma$-ray flux in the $n$th energy bin attributed to the Sgr dSph, and also yields an uncertainty $\sigma^{\rm def}_{\mathcal{N},n}$ on this quantity. This represents the statistical error arising from measurement uncertainties. However, there are also systematic uncertainties stemming from our imperfect knowledge of the templates characterising the other emission sources. To estimate these, we examine the five alternative models listed in \autoref{tab:loglikelihood} as ``Alternative background templates'', where we use different templates for the hadronic plus bremsstrahlung and inverse Compton backgrounds. Each of these models $m$ also returns a central value $\mathcal{N}_n^m$ and an uncertainty $\sigma_{\mathcal{N},n}^m$ on the Sgr dSph flux. We use the uncertainty-weighted dispersion of these models as an estimate of the systematic uncertainty (e.g.,ref.~\cite{Ackermann2018}):
\begin{equation}
    \delta\mathcal{N}_n = \sqrt{\frac{1}{\sum_m \left(\sigma_{\mathcal{N},n}^m\right)^{-2}} \sum_m \left(\sigma_{\mathcal{N},n}^m\right)^{-2} \left(\mathcal{N}_n^{\rm def} - \mathcal{N}^m_n\right)^2},
\end{equation}
where the sums run over the $m=6-1$ alternative models. We take the total uncertainty on the Sgr dSph flux in each energy bin to be a quadrature sum of the systematic and statistical uncertainties, i.e., $(\sigma^{\rm def,tot}_{\mathcal{N},n})^2 = (\sigma^{\rm def}_{\mathcal{N},n})^2 + \delta\mathcal{N}_n^2$. We plot the central values and uncertainties of the fluxes for the default model derived in this manner in \autoref{fig:luminosities}.

We have carried out several validation tests of this pipeline, which we describe in the Supplementary Information (SI).

\subsection*{Spectral modelling}

We model the observed Sgr dSph $\gamma$-ray spectrum as a combination of prompt magnetospheric MSP emission and IC emission from \pairs~escaping MSP magnetospheres. We construct this model as follows. The prompt component is due to curvature radiation from \pairs~in within MSP magnetospheres. 
The \pairs~energy distribution can be approximated as an exponentially-truncated power law \cite{Abdo2013,Song2021}
\begin{equation}
    \frac{dN_{\mathrm{MSP},e^\pm}}{dE_{e^\pm}} \propto E_{e^\pm}^{\gamma_\mathrm{MSP}} \exp\left(-\frac{E_{e^\pm}}{E_{\mathrm{cut},e^\pm}}\right),
\label{eq:promptSpec}
\end{equation}
and curvature radiation from these particles has a rate of photon emission per unit energy per unit time
\begin{equation}
    \frac{d\dot{N}_{\rm \gamma,prompt}}{dE_\gamma} = \mathcal{N}\left(L_{\gamma,\mathrm{prompt}}\right) E_\gamma^{\alpha} \exp\left(-\frac{E_\gamma}{E_{\rm cut, prompt}} \right),
\label{eq:cutoffpwrlaw}
\end{equation}
where $E_\gamma$ is the photon energy, $\mathcal{N}(L_{\gamma,\mathrm{prompt}})$ is a normalisation factor chosen so that the prompt component has total luminosity $L_{\gamma,\mathrm{prompt}}$, the index $\alpha$ is related to that of the \pairs~distribution by $\alpha = (\gamma_{\rm MSP} - 1)/3$, and the photon cutoff energy is related to the \pairs~cutoff energy by \cite{Baring2011}
\begin{equation}
E_{\rm cut,prompt} = \frac{3 \hbar c}{2 \rho_c}   \left(\frac{E_{\rm cut,e^\pm}}{m_e}\right)^3 \simeq 2.0 \ {\rm GeV} \ \left(\frac{\rho_c}{\rm 30 \ km}\right)^{-1} \left(\frac{E_{\rm cut,e^\pm}}{\rm 3 \ TeV}\right)^3
\label{eq:EcutPrompt}
\end{equation}
where $m_e$ is the electron mass, $\rho_c$ is the radius of curvature of the magnetic field lines, and the other symbols have the usual meanings. Given the rather small magnetospheres, we expect $\rho_c$ to be a small multiple of the $\sim$ 10 km neutron star characteristic radius; henceforth we set $\rho_c =$ 30 km. Empirically, $L_{\gamma,\rm prompt}$ is $\sim 10\%$ of the total MSP spin-down power \cite{Abdo2013}.

A larger proportion of the spin-down power goes into a wind of \pairs~escaping the magnetosphere. In the ultra-low density environment of the Sgr dSph, ionization and bremmstrahlung losses for this population, which occur at a rate proportional to the gas density, are negligible. Synchrotron losses, which scale as the magnetic energy density, will also be negligible; as noted in the main text, observed magnetic fields in dwarf galaxies are very weak \cite{Regis2015}, and we can also set a firm upper limit on the Sgr dSph magnetic field strength simply by noting that the magnetic pressure cannot exceed the gravitational pressure provided by the stars since, if it did, that magnetic field, and the gas to which it is attached, would blow out of the galaxy in a dynamical time. The gravitational pressure is $P \approx (\pi/2) G \Sigma^2$, where $\Sigma = M/\pi R^2$ is the surface density, and using our fiducial numbers $M = 10^8$ M$_{\odot}$ and $R = 2.6$ kpc gives an upper limit on the magnetic energy density 0.06 eV / cm$^3$; non-zero gas or cosmic ray pressure would lower this estimate even further. This is a factor of four smaller than the energy density of the CMB, implying that synchrotron losses are at most a 20\% effect, and can therefore be neglected.

This analysis implies that the only significant loss mechanism for these \pairs~is IC emission, resulting in a steady-state \pairs~energy distribution
\begin{equation}
\frac{dN_{e^\pm}}{dE_{e^\pm}} \propto E_{e^\pm}^\gamma \exp\left(-\frac{E_{e^\pm}}{E_{\mathrm{cut},e^\pm}}\right),
\end{equation}
where $\gamma = \gamma_{\rm MSP}-1$. We compute the IC photon distribution produced by these particles following ref.~\cite{Khangulyan2014}, assuming that ISRF of the \sgr is the sum of the CMB 
and two subdominant contributions, one consisting of light escaping from the Milky Way and the other a dilute stellar blackbody radiation field due to the stars of the dwarf.
We estimate the Milky Way contribution to the photon field at position of the dwarf using GalProp \cite{Porter:2017vaa},
which predicts a total energy density of $0.095$ eV/cm$^3$ (compared to 0.26 eV/cm$^{-3}$ for the CMB), comprised of 5 dilute black bodies
with colour temperatures and dilution factors
$\{ T_{\rm rad},\kappa \}$ as follows: 
$\{40 \ {\rm K }, 1.4 \times 10^{-6} \},
\{430 \ {\rm K }, 3.0 \times 10^{-11} \},
\{3400 \ {\rm K }, 4.3 \times 10^{-14} \},
\{6400 \ {\rm K }, 4.0 \times 10^{-15} \},$
and
$ \{26000 \ {\rm K }, 8.0 \times 10^{-18} \}
$.
We characterise the intrinsic light field of the dwarf as having a
colour temperature 3500 K and dilution factor of $7.0\times 10^{-15}$ (giving energy density $0.005$ eV cm$^{-3}$; these choices are those expected for a spherical region of radius 2.6 kpc and stellar luminosity $2\times 10^8$ $L_\odot$, the approximate parameters of the Sgr dSph).
This yields an IC spectrum
\begin{equation}
    \frac{d\dot{N}_{\gamma,\mathrm{IC}}}{dE_\gamma} = \mathcal{N}\left(L_{\gamma,\rm IC}\right) F\left(\gamma,E_{\mathrm{cut},e^\pm}\right),
\end{equation}
where $\mathcal{N}\left(L_{\gamma,\rm IC}\right)$ is again a normalisation chosen to ensure that the total IC luminosity is $L_{\gamma,\rm IC}$, and $F\left(\gamma,E_{\mathrm{cut},e^\pm}\right)$ is the functional form given by equation 14 of ref.~\cite{Khangulyan2014}, which depends on the \pairs~spectral index $\gamma$ and cutoff energy $E_{\mathrm{cut},e^\pm}$.

Combining the prompt and IC components, we may therefore write the complete emission spectrum as
\begin{equation}
    \frac{d\dot{N}_\gamma}{dE_\gamma} = \mathcal{N}\left(L_{\gamma,\mathrm{prompt}}\right) E_\gamma^{\alpha} \exp\left(-\frac{E_\gamma}{E_{\rm cut, prompt}} \right) + \mathcal{N}\left(L_{\gamma,\rm IC}\right) F\left(\gamma,E_{\mathrm{cut},e^\pm}\right).
\end{equation}
This model is characterised by four free parameters: the total prompt plus IC luminosity $L_{\gamma,\rm tot} = L_{\gamma,\rm prompt} + L_{\gamma,\rm IC}$, the ratio of the prompt and IC luminosities $f = L_{\gamma,\rm prompt}/L_{\gamma,\rm IC}$, the spectral index $\alpha$ of the prompt component (which in turn fixes the other two spectral indices $\gamma_{\rm MSP}$ and $\gamma$), and the cutoff energy for the prompt component $E_{\rm cut, prompt}$ (which then fixes the \pairs~cutoff energy $E_{\mathrm{cut},e^\pm}$).
Note that we make the simplest assumption that $\alpha$ and $E_{\rm cut, prompt}$ are uniform across the MSP population.
In reality, there may be a distribution of these properties but the parameteric form of \autoref{eq:promptSpec}
provides a good description, in general, 
of both individual MSP spectra and the
aggregate spectra of GC MSP populations \cite{Song2021}.

We fit the observed \sgr~spectrum to this model using a standard $\chi^2$ minimisation, using the combined statistical plus systematic uncertainty.
We obtain an excellent fit: the minimum $\chi^2$ is 7.7 for 15 (data points) - 4 (fit parameters) = 11 (degrees of freedom, dof) or a reduced $\chi^2$ of 0.70.
We report the best-fitting parameters in Supplementary~\autoref{tab:table1}, and plot the result best-fit spectra over the data in \autoref{fig:luminosities}; we show the best-fit estimate (with $\pm 1\sigma$ confidence region) for the magnetospheric luminosity per stellar mass of the \sgr MSPs in \autoref{fig:LgammaOvrMstar}. 

We also carry out an additional consistency check, by comparing our best-fit parameters describing the prompt emission -- $\alpha$ and $E_{\rm cut,prompt}$ -- to direct measurements of the prompt component from nearby, resolved MSPs \cite{Abdo2013,Song2021}, and to measurements of GCs, whose emission is likely dominated by unresolved MSPs \cite{Song2021}. We carry out this comparison in Supplementary~\autoref{fig:FitContours}. In this figure, we show joint confidence intervals on $\alpha$ and $E_{\rm cut,prompt}$ from our fit. For comparison, we construct confidence intervals for $\alpha$ and $E_{\rm cut,prompt}$ from observations using the sample of ref.~\cite{Song2021}, who fit the prompt emission from 40 GCs and 110 individually-resolved MSPs. We draw 100,000 Monte Carlo samples from these fits, treating the stated uncertainties as Gaussian, and construct contours in the $(E_{\rm cut,prompt}, \alpha)$ plane containing 68\%, 95\%, and 99\% of the sample points. As the plot shows, the confidence region from our fit is fully consistent with the confidence regions from the observations, indicating that our best-fit parameters are fully consistent with those typically observed for MSPs and GCs.

\section*{Data availability}

All data analysed for this study are publicly available.
In particular, \fermi \ data are available from \url{https://fermi.gsfc.nasa.gov/ssc/data/} and Gaia data are available from \url{https://gea.esac.esa.int/archive/}.
The statistical pipeline, astrophysical templates, and gamma-ray observations necessary to reproduce our main results are publicly available in the following zenodo repository: \url{10.5281/zenodo.6210967}.

\section*{Code availability}

\fermi \ data used in our study were reduced and analysed  using the standard  
\textsc{Fermitools v1.0.1}
software package available from \url{https://github.com/fermi-lat/Fermitools-conda/wiki}.
The performance of the \fermi \ was modelled with the \texttt{P8R3\_ULTRACLEANVETO\_V2} Instrument Response Functions (IRFs).
Spectral analysis and fitting was performed using custom \textsc{MATHEMATICA}
code created by the authors which is available upon reasonable request.


\section*{Acknowledgements}

RMC acknowledges 
support from the Australian Government through the Australian Research Council, award
DP190101258 (shared with MRK)
and hospitality from the Virginia Institute of Technology, the Max-Planck Institut f\"ur Kernphysik, and the GRAPPA Institute at the University of Amsterdam supported by the Kavli IPMU at the University of Tokyo. 
O.M. is supported by the GRAPPA Prize Fellowship and JSPS KAKENHI Grant Numbers JP17H04836, JP18H04340, JP18H04578, and JP20K14463.
This work was supported by World Premier International Research Centre Initiative (WPI Initiative), MEXT, Japan. 
ADM acknowledges support from the Australian Government through a Future Fellowship from the Australian Research Council, award FT160100206.
M.R.K. acknowledges support from the Australian Government through the Australian Research Council, award
DP190101258 (shared with RMC) and
FT180100375.
The work of S.A. was supported by MEXT KAKENHI Grant Numbers JP20H05850 and JP20H05861.
The work of S.H.\ is supported by the U.S.\ Department of Energy Office of Science under award number DE-SC0020262 and NSF Grant No.\ AST-1908960 and No.\ PHY-1914409. The work of DS is supported by the U.S.\ Department of Energy Office of Science under award number DE-SC0020262. 
T.V. and A.R.D. acknowledge the support of the Australian Research Council's Centre of Excellence for Dark Matter Particle Physics (CDM) CE200100008.
AJR acknowledges support from the Australian Government through the Australian Research Council, award FT170100243.
RMC thanks Elly Berkhuijsen, Rainer Beck, Ron Ekers, Matt Roth, and Thomas Siegert for useful communications.

\section*{Author contributions statement}

R.M.C. initiated the project and led the spectral analysis and theoretical interpretation. 
O.M constructed the astrophysical templates, designed the analysis pipeline, and performed the data analysis of  $\gamma$ ray observations.
D.M., M.R.K., C.G., R.J.T., F.A., J.A.H., S.A., S.H., A.G., M.R., L.F., and A.R. provided theoretical insights and interpretation, and advice about statistical analysis.
T.V. and A.R.D. provided insights on the expected distribution of dark matter.
R-Z.Y. performed an initial \gr \ data analysis.
M.D.F helped with radio data.
The main text was written by RMC, MRK, and O.M. and the Methods section was written by O.M., R.M.C., and MRK.
All authors were involved in the interpretation of the results and all reviewed the manuscript.
%



\subsection*{Competing interests}

The authors declare no competing interests.



\begin{table}
    \centering
    \small
    \begin{tabular}{llll@{\qquad\qquad}rrrr}
    \hline\hline
    \multicolumn{4}{c}{Template choices} & \multicolumn{4}{c}{Results} \\
    Hadr. / Bremss. & IC & FB & Sgr dSph &
    $-\log(\mathcal{L}_{\rm Base})$  & $-\log(\mathcal{L}_{{\rm Base}+{\rm  Sgr}})$  &  $\mbox{TS}_{\rm Source}$& Significance \\[0.5ex] \hline 
    \multicolumn{8}{c}{Default model} \\[0.5ex]
    HD & 3D & S & Model I & 866680.6 &866633.0  & 95.2  &  $8.1\;\sigma$ \\[0.5ex] \hline
    \multicolumn{8}{c}{Alternative background templates} \\[0.5ex]
    HD & 2D A & S & Model I   &  866847.1    & 866810.9  & 72.3  &  $6.9\;\sigma$ \\
    HD & 2D B & S & Model I &  867234.9  &867192.1  & 85.8 &  $7.8\;\sigma$ \\
    HD & 2D C & S & Model I &  866909.4  & 866868.5  & 81.7  & $7.4\;\sigma$ \\
    Interpolated & 3D & S & Model I   & 867595.4     & 867567.4  & 56.0 & $5.8\;\sigma$ \\
    GALPROP & 3D & S & Model I & 866690.5   &  866640.8 & 99.5  & $8.3\;\sigma$ \\[0.5ex] \hline
    \multicolumn{8}{c}{Flat FB template} \\[0.5ex]
    HD & 3D & U & Model I & 867271.7 & 867060.1 & 423.2 &  $19.1\;\sigma$ \\
    HD & 2D A & U & Model I & 867284.2 &867122.9 & 322.5 &  $16.5\;\sigma$ \\
    HD & 2D B & U & Model I & 867624.3 & 867464.0& 320.7 & $16.4\;\sigma$ \\
    HD & 2D C & U & Model I & 867322.7 &867158.2 &329.0 & $16.6\;\sigma$ \\
    Interpolated & 3D & U & Model I & 867287.4 & 867081.2& 412.4 & $18.9\;\sigma$ \\
    GALPROP & 3D & U & Model I & 868214.6 & 868040.9& 347.6  & $17.2\;\sigma$ \\[0.5ex]\hline
    \multicolumn{8}{c}{Alternative Sgr dSph templates} \\[0.5ex]
    HD & 3D & S & Model II & 866680.6 & 866626.3 & 108.5  & $8.7\;\sigma$ \\
    HD & 3D & S & Model III & 866680.6 & 866647.5 & 66.1  & $6.4\;\sigma$ \\
    HD & 3D & S & Model IV & 866680.6 & 866678.2 & 4.8 & $0.4\;\sigma$ \\
    HD & 3D & S & Model V & 866680.6 &866644.9 & 71.5  & $6.7\;\sigma$ \\
    HD & 3D & U & Model II & 867271.7 & 866970.7  & 602.1  & $23.2\;\sigma$ \\
    HD & 3D & U & Model III & 867271.7 & 866994.1  & 555.3 & $22.2\;\sigma$ \\
    HD & 3D & U & Model IV & 867271.7 & 867152.2 & 239.1 & $14.0\;\sigma$ \\
    HD & 3D & U & Model V & 867271.7 & 866993.3 & 556.9   & $22.2\;\sigma$ \\
    \hline\hline
    \end{tabular}
    \caption{{\bf Template analysis results comparing {\it baseline} to {\it baseline + Sgr dSph} models}. 
    Columns (1) - (3) specify the {baseline} templates used for Galactic hadronic / bremsstrahlung emission, inverse Compton emission, and the Fermi Bubbles, respectively.
    Column (4) specifies {source} templates
    describing the Sgr dSph (see Methods for details). Columns (5) and (6) give the log likelihood for the baseline model (without the Sgr dSph) and the baseline + Sgr dSph model, and columns (7) and (8) give the test statistic with which the baseline + Sgr dSph model is preferred, and the corresponding statistical significance of that preference. 
    The improvement in TS going from $\{\rm HD, 3D, U , Model \ I\}$ to $\{\rm HD, 3D, S , Model \ I\}$ is $\Delta$TS = 854.2, equivalent to 28.0 $\sigma$.
    Note that Sgr dSph model IV -- which generates a statistically insignificant improvement to the baseline for one particular combination in the last cluster -- is the sparsest stellar template, containing only 675 stars.
    }
    \label{tab:loglikelihood}
\end{table}

\newpage

\section*{Figures}

\begin{figure}[H]
    \centering
\includegraphics[width=\linewidth]{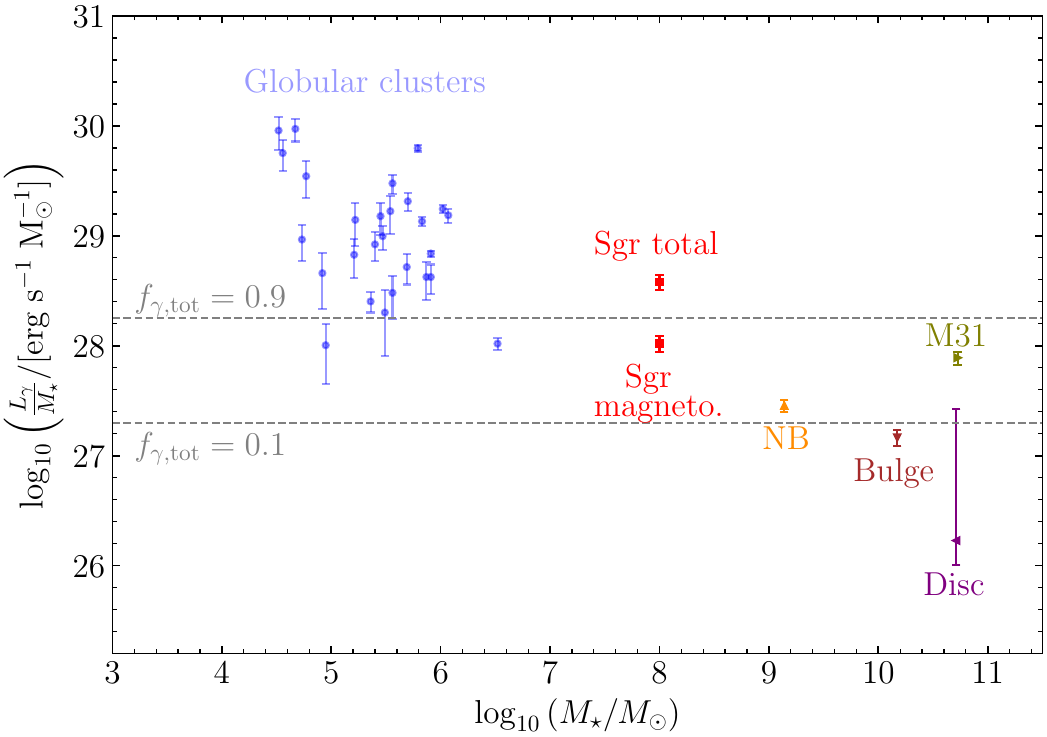}
\caption{{\bf \gr \ luminosity
normalised to stellar mass for various structures whose emission is plausibly dominated by MSPs.}
The `Sgr magneto.' datum shows our
best-fit magnetospheric luminosity per stellar mass
(the spectrum shown as the dashed blue curve in \autoref{fig:luminosities})
while the `Sgr tot' datum is the total measured luminosity
associated to the \sgr template. Globular cluster (`GC') measurements are from ref.~\cite{Song2021}, while the remaining data (collated by ref.~\cite{Song2021}) are from ref.~\cite{Macias2019} (nuclear bulge of the Milky Way, `NB'), ref.~\cite{Ackermann2017} (M31), and ref.~\cite{Bartels2018} (Milky Way disc). 
Error bars show 1$\sigma$ errors.
The horizontal, dashed, grey curves show
the predicted total 
\gr \ luminosity per unit stellar mass
at the nominated efficiencies, $f_{\rm \gamma,tot} = \{0.1, 0.9\}$, 
given an MSP spin-down power per unit stellar mass of
$2 \times 10^{28}$   erg/s/$\msun$ as
we infer from ref.~\cite{Sudoh2020}.}
\label{fig:LgammaOvrMstar}
\end{figure}

\newpage

\printbibliography[segment=\therefsegment,check=onlynew]

\clearpage

\newrefsegment

\section*{Extended Data}
\setcounter{figure}{0}
\setcounter{table}{0}

\renewcommand{\figurename}{Extended Data Figure}
\renewcommand{\tablename}{Extended Data Table}

\begin{figure*}[ht!]
\centering
\includegraphics[width=0.6\textwidth,angle =270]{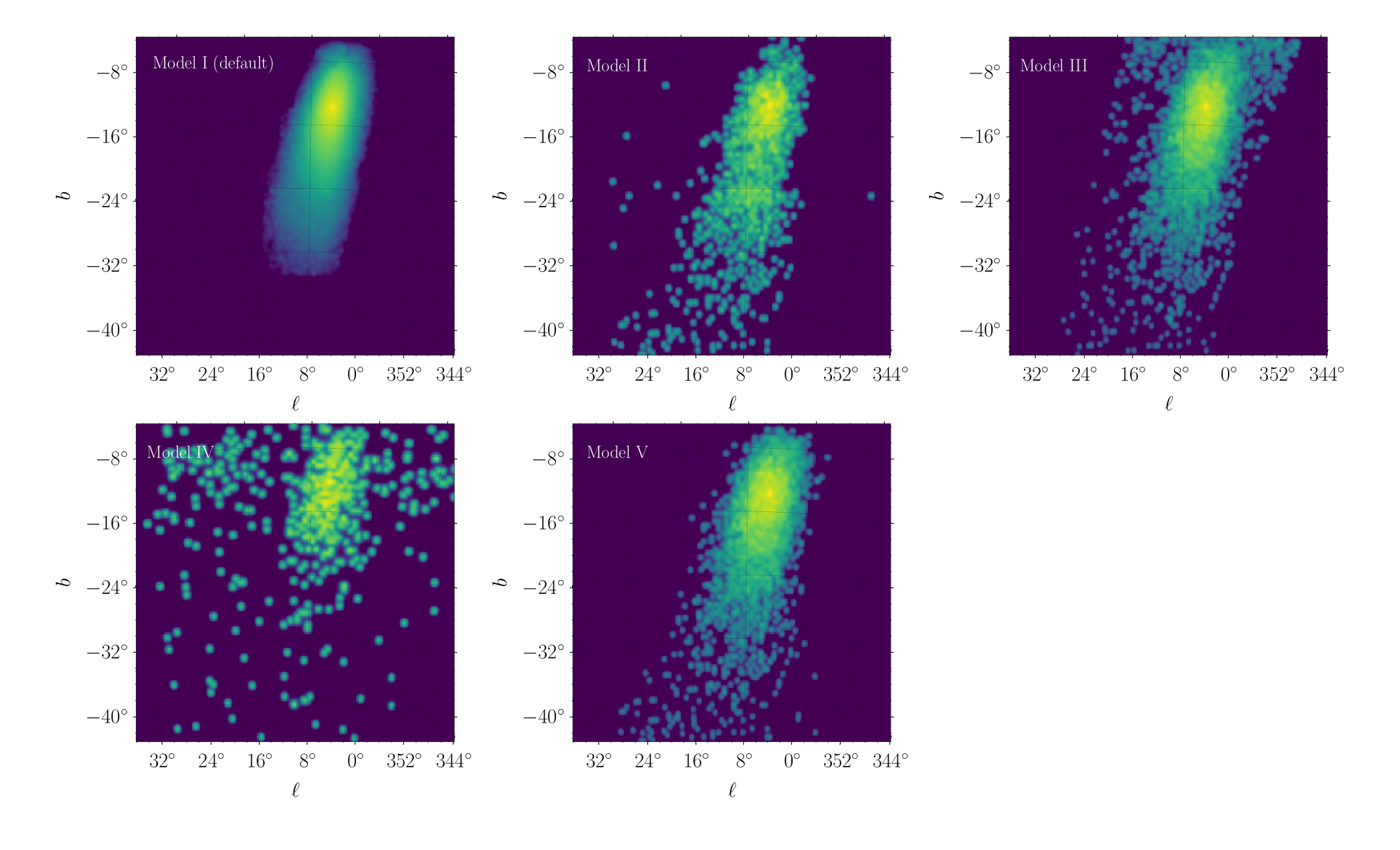}
\caption{The stellar density templates for the Sgr dSph used in this study. Each map has been normalized, so the units are arbitrary; the color scale is logarithmic. Morphological differences among the templates are due to different stellar candidates (red clump or RR Lyrae), search algorithms, and search target (the dwarf remnant or the stream).  
Data sources are as follows: 
Model I, ref.~\cite{Vasiliev2020};
Model II, ref.~\cite{Ibataetal:2020};
Model III, ref.~\cite{Iorio2019};
Model IV and Model V, ref.~\cite{Ramosetal:2020}.
Detailed descriptions of these templates are given in the S.I.~sec.~2.}
\label{fig:Stellartemplates}
\end{figure*}

\begin{figure}[t!]
\centering
\includegraphics[scale=0.3]{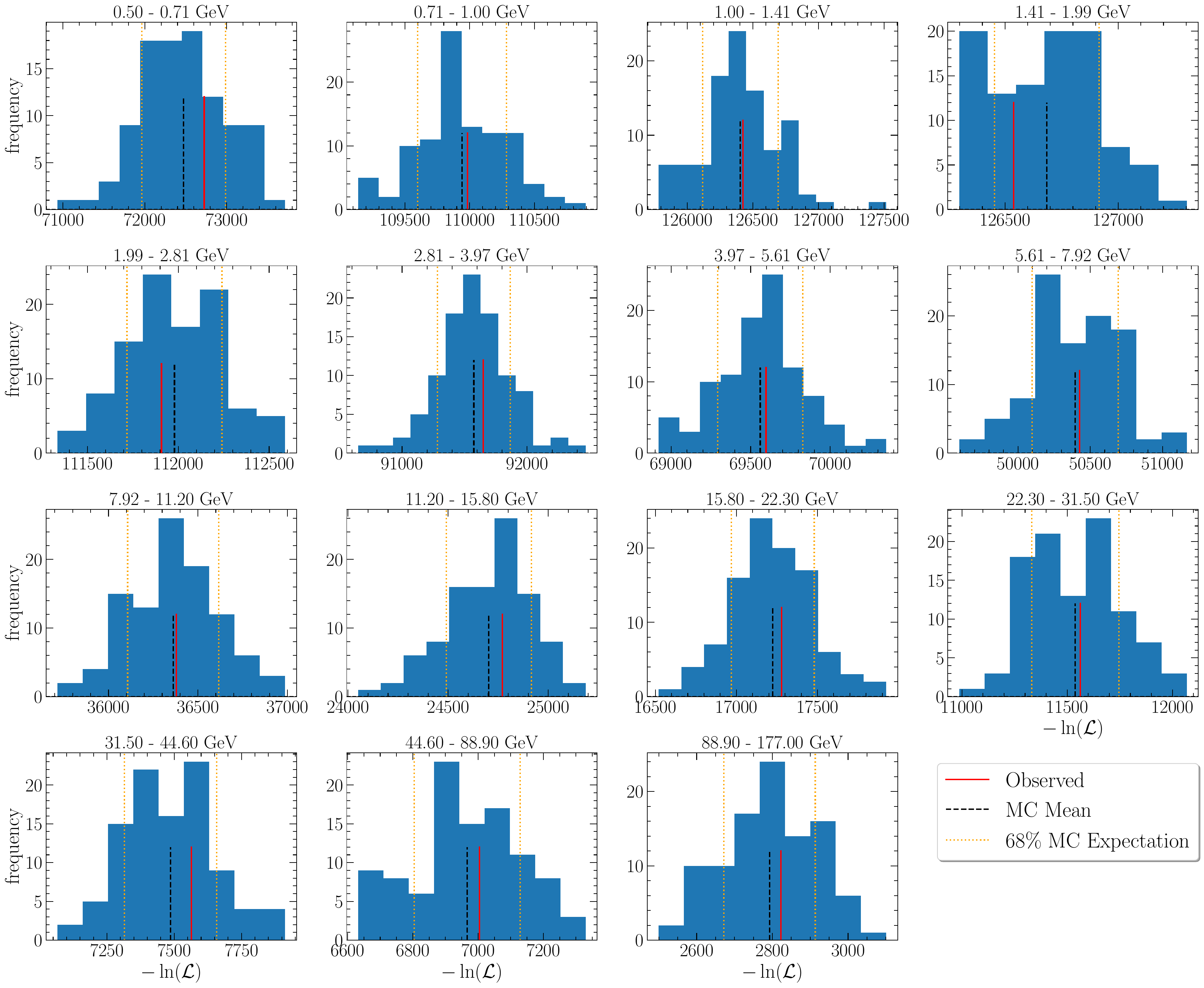} \\
\caption{Goodness of fit computation for the best-fitting baseline + Sgr dSph model using our preferred set of templates (first entry in \autoref{tab:loglikelihood}). In each of the 15 panels, one for each of the energy bins in our analysis pipeline, the blue histograms show the distribution of $-\ln\mathcal{L}$ values produced in 100 Monte Carlo trials where we use our pipeline to fit a mock data set produced by drawing photons from the same set of templates used in the fit; orange dashed vertical lines show the 68\% confidence range of this distribution, and black dashed vertical lines show the mean. 
Under the 
hypothesis that our best-fitting model for the real \textit{Fermi} observations is a true representation of the data, and that disagreements between the model and the data are solely the result of photon counting statistics, the log-likelihood values for our best-fitting model should be drawn from the distributions shown by the blue histograms. For comparison, the red vertical line shows the actual measured log likelihoods for our best fit. The fact that these measured values are well within the range spanned by the Monte Carlo trials
indicates that we cannot rule out this 
 hypothesis, indicating that our model is as good a fit to the data as could be expected given the finite number of photons that \textit{Fermi} has observed.
}\label{fig:fitvalidation}
\end{figure}

\begin{figure*}[ht!]
\centering
\includegraphics[width=1.0\textwidth]{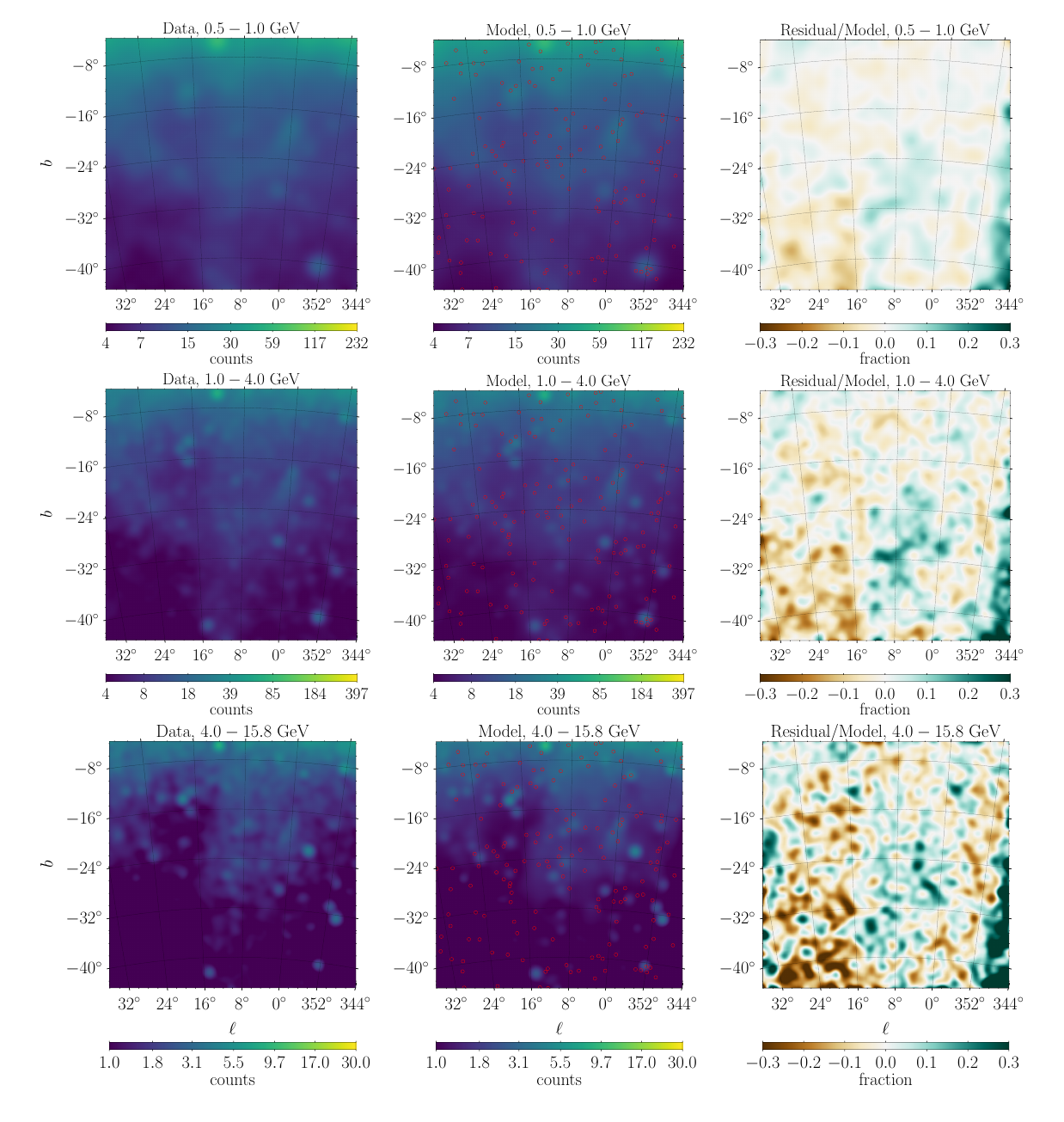}
\caption{Measured photon counts (left), best-fit baseline + \sgr model (middle), and the fractional residuals $(Data-Model)/Model$ (right). The images were constructed by summing the corresponding energy bins over the energy ranges displayed on top of each panel: [0.5, 1.0] GeV, [1.0, 4.0] GeV, [4.0, 15.8] GeV, from top to bottom. The maps have been smoothed with Gaussian filters of radii $1.0^\circ$, $0.8^\circ$, and $0.5^\circ$ for each energy range displayed, respectively
(where these angular scales are determined by the \fermi \ point spread function at the low-edge of the energy interval for the former two, while the latter is determined by the angular resolution of the gas maps).
The spectrum of baseline + \sgr model components shown here can be seen in ~\autoref{fig:totalspectra}. The 4FGL~\cite{Fermi-LAT:4FGL} \gr \ point sources included in the baseline model are represented by the red circles.
}
\label{fig:Residuals}
\end{figure*}

\begin{figure}
\centering
\includegraphics[scale=0.9]{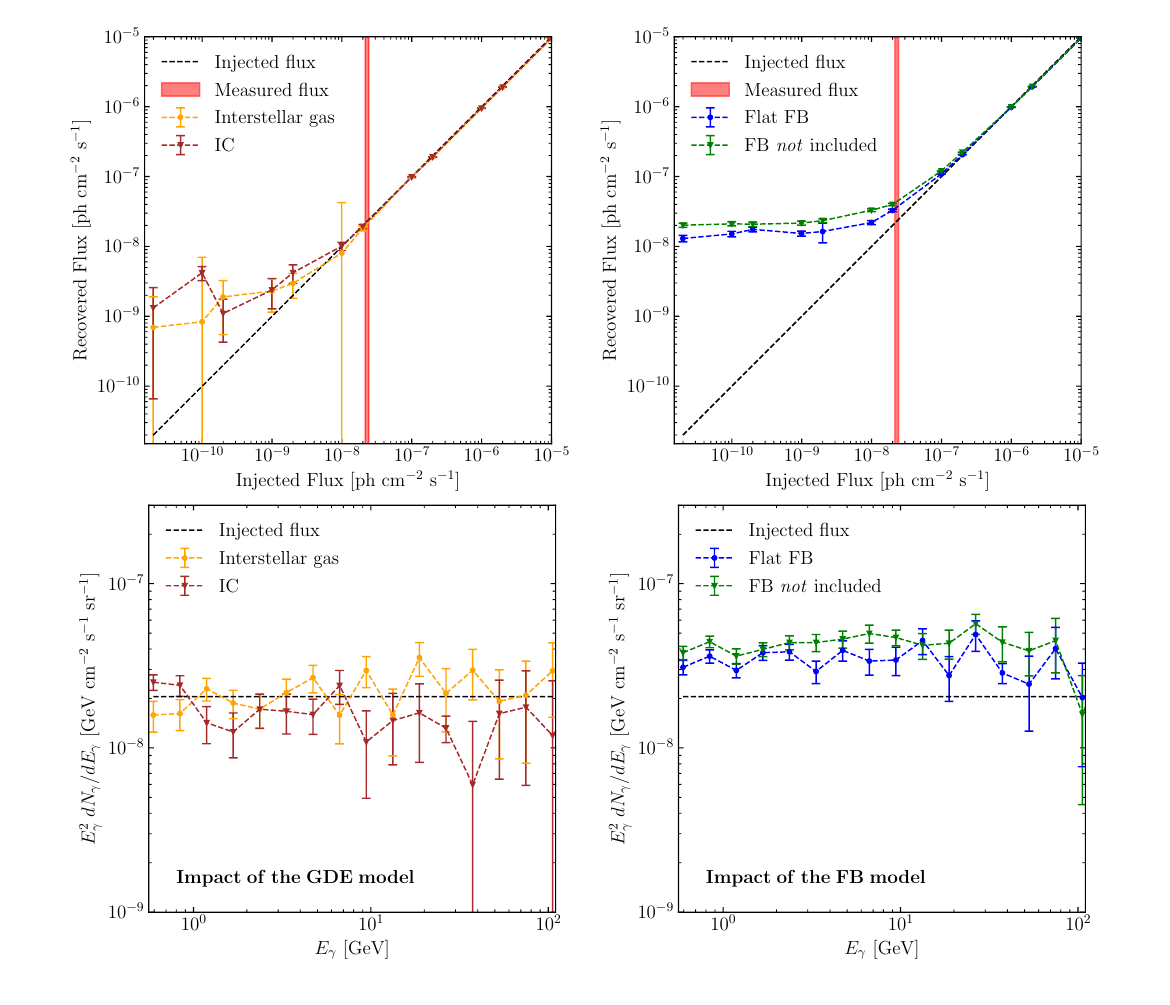}
\caption{
Results from our template mismatch tests. Each of the coloured lines shows the results of a test where we generate synthetic data with one set of templates, and attempt to recover the Sgr dSph in those data using a different set. In the upper two panels, the horizontal axis shows the true, energy-integrated Sgr dSph photon flux in the synthetic data, while the vertical axis shows the value (with $1 \sigma$ statistical error bars) retrieved by our pipeline; the black dashed lines indicate perfect recovery of the input, and the vertical bands show the photon flux we measure for the Sgr dSph in the real \textit{Fermi} data. In the bottom two panels we plot the recovered energy flux in each energy bin (with $1 \sigma$ statistical error bars), for the case where the injected photon flux most closely matches the real Sgr dSph flux; the black dashed line again shows perfect recovery of the injected signal. The left panels show experiments where we mismatch the Galactic hadronic and IC templates, while the right panels show experiments where we mismatch the FB templates; see Methods for details.
}\label{fig:injectionrecovery}
\end{figure}

\begin{figure}
\centering
\includegraphics[scale=0.9]{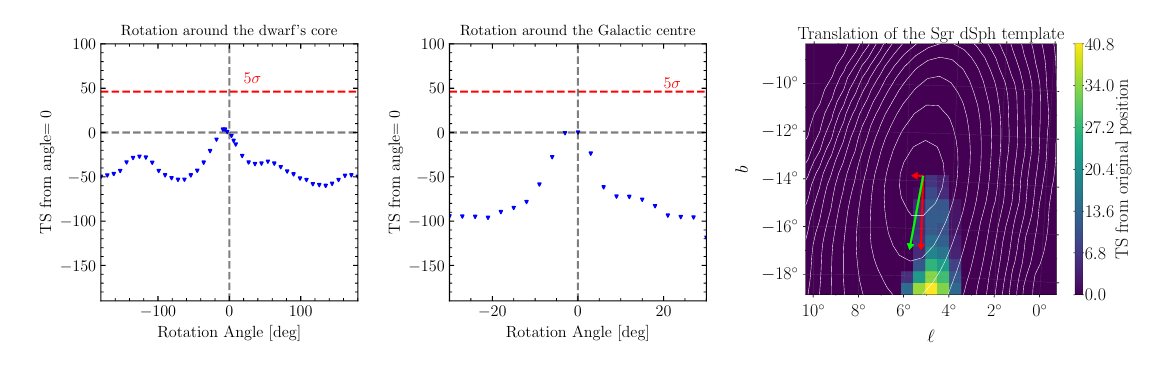}
\caption{
Results of our rotation and translation tests. \textit{Left:} change in TS when repeating the analysis using the default baseline + Sgr dSph model, but with the Sgr dSph rotated about its centre by the indicated angle (blue points); TS values $>0$ indicate an improved fit (dashed grey line), with $\mbox{TS} = 46.1$ corresponding to a $5\sigma$-significant improvement (red dashed line). 
\textit{Centre:} same as the left panel, but for tests with the Sgr dSph template rotated about the Milky Way centre, rather than its own centre. \textit{Right:} tests for translation of the Sgr dSph template. The true position of the Sgr dSph centre is the center of the plot, and the colour in each pixel indicates the change in TS if we displace the Sgr dSph centre to the indicated position; the maximum shown, at a displacement $\Delta b \approx -4^\circ$, has $\mbox{TS} = 40.8$, corresponding to $4.5\sigma$ significance. For comparison, white contours show the original, unshifted Sgr dSph template, and the green arrow shows the direction anti-parallel to the Sgr dSph's proper motion, back along its past trajectory; red arrows show the projection of the green arrow in the $\ell$ and $b$ directions.
}\label{fig:rotationAndTranslationTests}    
\end{figure}

\begin{figure}
\centering
\includegraphics[scale=1]{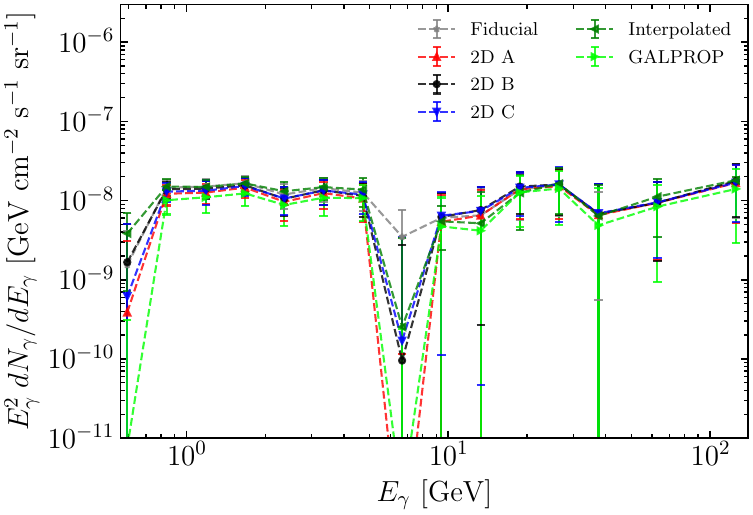}\\
\caption{\sgr spectra derived from template analysis using different Galactic diffuse emission models; in all cases the spectrum shown is the flux averaged over the entire ROI, not the flux within the footprint of the Sgr dSph template. The fiducial model is our default choice (first entry in Table~\ref{tab:loglikelihood}), while other lines correspond to alternate foregrounds -- models 2D A (red), 2D B (black), and 2D C (blue) for the Galactic IC foreground, and models Interpolated (dark green) and GALPROP 3D-gas (light green) for the Galactic hadronic + bremsstrahlung foreground. The error bars display $1\sigma$ statistical errors. See Table~\ref{tab:loglikelihood} and text for details.}
\label{fig:SgrSpecVar}
\end{figure}

\begin{figure}
\centering
\includegraphics[scale=1]{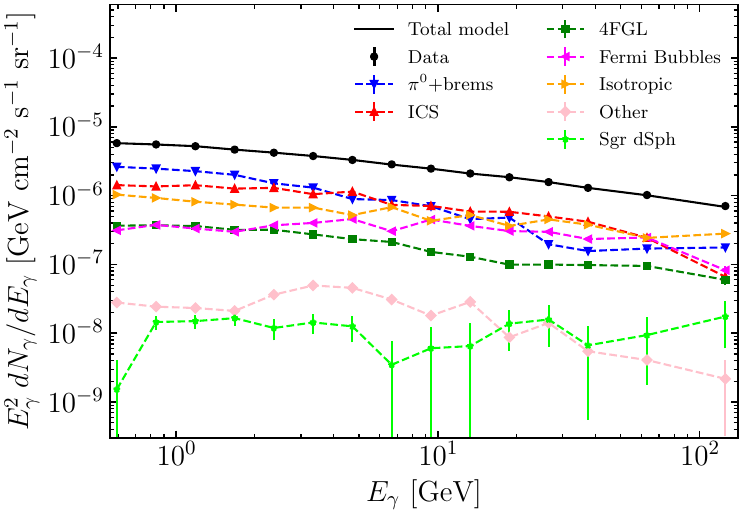}\\ 
\caption{
Contribution of each template component to the $\gamma$-ray spectrum averaged over the entire ROI, for our default baseline + \sgr model. Components shown are as follows: $\pi^0+\mbox{brems}$ is the Galactic hadronic plus bremsstrahlung foreground, ICS is the Galactic inverse Compton foreground, 4FGL indicates point sources from the 4th \textit{Fermi} catalogue, Fermi Bubbles indicates the structured Fermi Bubble template, isotropic is the isotropic $\gamma$-ray background, ``other'' includes the Sun and Moon, Loop I, and the Galactic Centre Excess, and Sgr stream indicates the Sgr dSph.
The error bars display $1\sigma$ statistical errors.
}
\label{fig:totalspectra}
\end{figure}



\clearpage

\newrefsegment

\renewcommand{\figurename}{Supplementary  Figure}
\renewcommand{\tablename}{Supplementary  Table}

\setcounter{figure}{0}
\setcounter{table}{0}

\section*{Supplementary Information}

\section{Chance overlap calculation}
\label{sec:overlap}

In the main text we estimate the probability of a chance overlap between cocoon \gr \ structure and \sgr to be $\approx 1\%$.
This follows simply from noting that the solid angle of the Bubbles is around 0.7 sr\cite{Ackermann2014} and the cocoon covers $\lesssim$20\% of this solid angle, so the chance probability for an overlap if these objects were placed randomly on the sky is $\lesssim 0.2 \times 0.7/(4 \pi) \sim 0.012$. However, this is a generous upper limit; it does not take into account that, as revealed by the template analysis, there is a much more detailed correspondence between the \gr \ substructure and the stellar distribution not accounted for here. Moreover, the naive 1\% estimate does include a `look-elsewhere' correction: the Milky Way is surrounded by satellite galaxies and there are apparently other regions of sub-structure within the Fermi Bubbles.
However, not only is the cocoon the brightest and first-discovered region of sub-structure \cite{Su2012}, it is also the only region that has been reliably detected by independent analyses \cite{Selig2015,Ackermann2014}, and is visibly-evident in 
independently-produced
\gr\ maps \cite{Yang2014,deBoer2015}.
The \sgr is also a special object: it is the
brightest MW satellite not yet (prior to this work) detected in $\gamma$-rays. 
(In fact, not only is the \sgr the brightest satellite undiscovered in $\gamma$-rays, it is substantially brighter than the next brightest galaxy\footnote{The list of all the MW satellites with apparent magnitude $m<10$ includes 8 objects,
the brightest two, the LMC and SMC, with $m \sim 0.3$ and $\sim 2.1$, respectively, are already detected in $\gamma$-rays.
The next brightest is the \sgr with  $m \sim 3$; after that come Fornax, Sculptor, and Leo I with $m \sim 7.3, 8.7$ and 10.0
and angular diameters of $0.24^\circ, 0.51^\circ$ and $0.11^\circ$, respectively, which, even assuming they could be detected, would at best only appear marginally extended to \fermi.}.)
Overall, we have a spatial overlap
(and detailed morphological correspondence as argued elsewhere) between 
the brightest region of substructure within the {\it Fermi} Bubbles
and the Sgr dSph, the
second closest, third-most massive, third brightest, and third most angularly extended satellite galaxy of the MW.

\section{Construction of the Sgr dSph templates}
\label{sec:stellarmapsDetails}

Here we provide detailed descriptions of 
how we construct the Sgr dSph templates shown in E.D.~\autoref{fig:Stellartemplates}.

\subsubsection*{Model I:} 

\begin{figure*}[t!]
    \centering
    \includegraphics[scale=0.15]{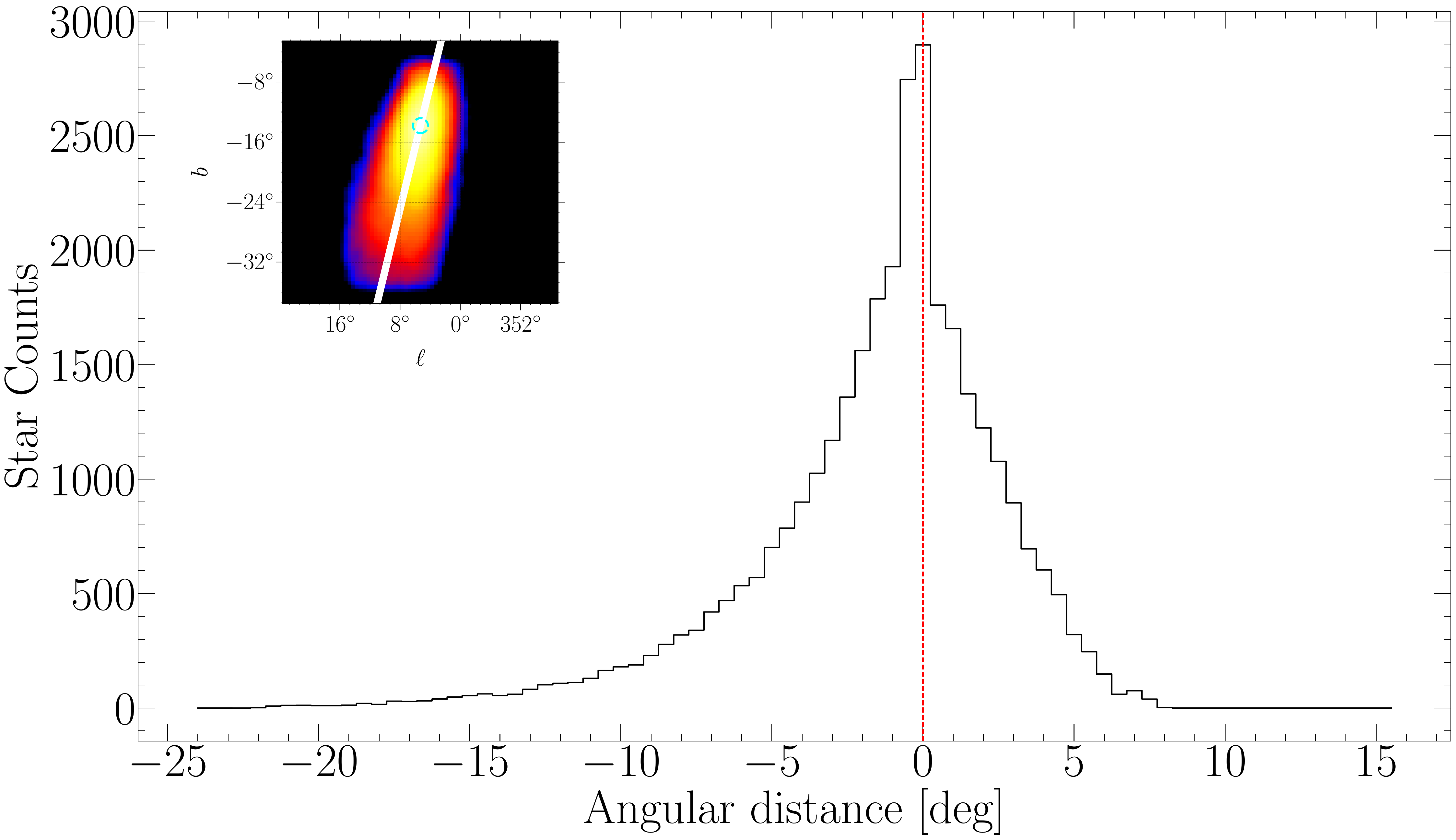}
    \includegraphics[scale=0.15]{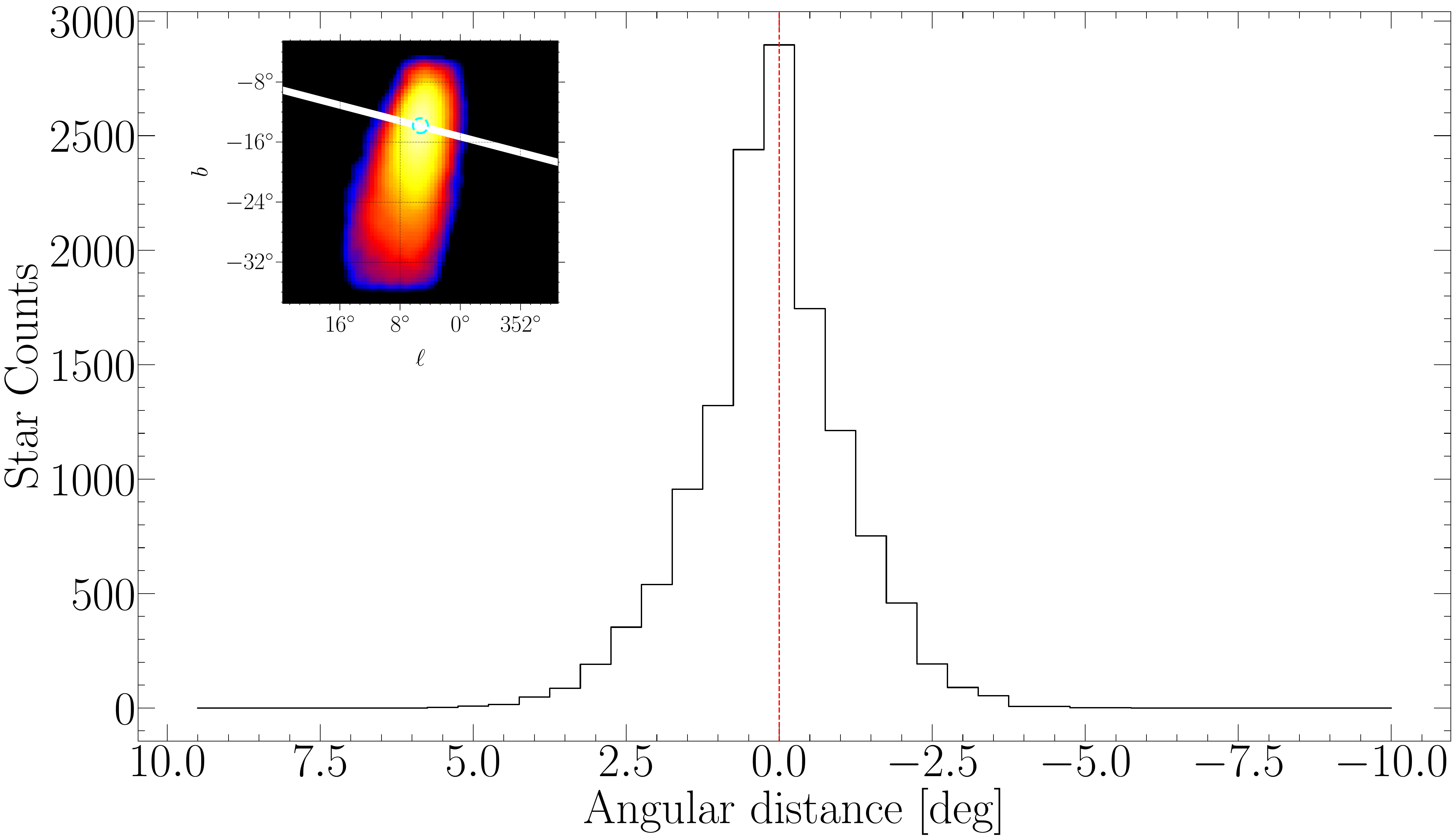}
    \caption{Star count profiles for Model I, showing number of stars measured in bins of angular distance from the gravitational centre of the dwarf along its long (top) and short (bottom) axes. In the inset images (which are identical to the first panel of \autoref{fig:Stellartemplates}), we mark the gravitational centre of the dwarf with a cyan circle, and show the long and short axes along which we measure the profiles as white bands.}
    \label{fig:profile_longaxis}
\end{figure*}

We extract this template from the stellar catalogue constructed in Ref.~\cite{Vasiliev2020}, which was derived using photometric and astrometric data from {\it Gaia} Data Release 2 (DR2), and kinematic measurements from various other surveys. The catalogue consists of a list of $2.26\times 10^5$ candidate member stars of the \sgr remnant, which are reliably separated from the field stars. Every object in the catalogue has an extinction-corrected G-band magnitude larger than 18, and more than half of the objects in this catalogue are classified as red clump stars. Note that Ref.~\cite{Vasiliev2020} adapted their procedure to reproduce the observed properties of the \sgr remnant, not the stream, which is why the first panel of E.D.~\autoref{fig:Stellartemplates} only shows the remnant. We show profiles of stellar number count along the long and short axes of the dwarf for this template in S.I.~\autoref{fig:profile_longaxis}. 

\subsubsection*{Model II:}

Our second template comes from ref.~\cite{Ibataetal:2020}. Instead of red clump stars, this study selected a sample of RR Lyrae stars from {\it Gaia} DR2 data, for which distances are accurately measured. Also, rather than focusing on member stars of the \sgr remnant, Ref.~\cite{Ibataetal:2020} used the \textsc{streamfinder} algorithm to single out stars with high probability of belonging to the Sagittarius Stream. By using the kinematic properties of the stars in that study, we constructed a template containing 2369 RR Lyrae stars (cf. \autoref{fig:Stellartemplates}) in our ROI. Note that the stellar number count in this map is approximately two orders of magnitude smaller than that in Model I.   

\subsubsection*{Model III:}

Ref.~\cite{Iorio2019} performed an all-sky analysis of RR Lyrae stars  (in {\it Gaia} DR2 data) belonging to globular clusters, dwarf spheroidal galaxies, streams, and the Magellanic Clouds. Our Model III template is a subset of their data identified as belonging to the Sgr dSph, selected to reproduce their Fig.~1 (bottom-right). It includes $1.31\times 10^4$ RR Lyrae stars in our ROI. 

\subsubsection*{Model IV and Model V:}

Ref.~\cite{Ramosetal:2020} developed two empirical catalogues of RR Lyrae stars in {\it Gaia} DR2 data, which form the basis for our final two templates. The first (Model IV), corresponds to the nGC3 sample, which is characterized for its lower-completeness and higher-purity. This template contains 675 stars in our ROI. The second (Model V), is the Strip sample, containing higher-completeness, but lower purity. The total number of stars in our ROI for this model is 4812.

\section{Validation tests}
\label{ssec:validation}

While our template analysis indicates a strong statistical preference for emission tracing the Sgr dSph, we also carry out five further validation tests to check the robustness of the result.

First, we check whether the residuals between the baseline + \sgr source model and the {\it Fermi} data from our ROI are consistent with the level expected simply as a result of photon counting statistics, using a method similar to that of Ref.~\cite{Buschmann2020}. Under the 
hypothesis that the {\it Fermi} data are a Poisson draw from our best-fit baseline + \sgr model (i.e., that our model is correct, and any differences between it and the actual data are simply due to shot noise), we can determine the expected distribution of $\ln\mathcal{L}_n$ values via Monte Carlo. For each Monte Carlo trial, we draw a set of mock photon counts $\Phi_{n,i,\rm mock}$ in each pixel and energy bin from our best-fitting model (multiplied by the instrument response function), and then compute the energy-dependent log-likelihood for this mock data set using the same pipeline we use on the real data. We repeat this procedure 100 times, and plot the distribution of log-likelihood values it produces 
as the blue histograms
in E.D. \autoref{fig:fitvalidation}. These
histograms represent the expected log likelihood in each energy bin under the 
hypothesis. We then compare this to the actual value of $\ln\mathcal{L}_n$ we measure for our model as compared to the real \textit{Fermi} data. The plot shows that our measured log-likelihood falls squarely within the range expected under the 
hypothesis, and we therefore conclude that the residuals between our model and the real data are consistent with being solely the result of photon counting statistics. 

In addition to testing whether the residuals between model and data are consistent with simply being shot noise when we sum over all pixels (which is what the likelihood measures), we can also examine the residuals as a function of position. We do so in E.D.~\autoref{fig:Residuals}, which shows the measured \textit{Fermi} counts in our ROI (summed in three energy bins) in the first column, our best-fitting baseline + \sgr model in the second column, and fractional residuals [$(\rm{Data}-\rm{Model})/\rm{Model}$] in the third column. The images are smoothed with a $0.5^\circ$ Gaussian kernel, since this is roughly the resolution of our interstellar gas maps~\cite{Macias2018,Macias2019}. The plot shows that, on a point-by-point basis, our models reproduce the data within $\sim 10\%$ over most of the ROI. 
There are, however, a few small patches of correlated residuals, which are only at the $\sim 30\%$ level, and are far from the Sgr dSph region. 
This points to the existence of real structure in the Fermi Bubbles that is not yet perfectly modelled, but given the small level of the residuals and the distance between them and the signal in which we are interested, 
we do not expect this modelling imperfection to significantly bias our results.

As our second validation test, we evaluate the sensitivity of our pipeline to uncertainties in our templates for Galactic diffuse emission, and we verify that our pipeline can recover synthetic signals similar to the Sgr dSph even when our templates are imperfect. Recall that we have three components of Galactic diffuse emission for which the templates are at least somewhat uncertain: hadronic + bremsstrahlung emission (for which our template can be HD, Interpolated, or GALPROP), Galactic IC emission (for which the template can be 3D, 2D A, 2D B, or 2D C), and the Fermi Bubbles (for which the template can be S, structured, or U, unstructured). 
We test the sensitivity of our fits to these template choices as follows. First, we generate a set of mock background data by drawing a random realisation of photon counts from one combination of these templates, and on top of this we add a synthetic Sgr dSph signal; the Sgr dSph photons follow the spatial morphology of our Sgr dSph model I template, have a spectral shape $dN_\gamma/dE_\gamma \propto E_\gamma^{-2}$, and have a normalisation that we vary systematically from $\approx 10^{-11}$ ph cm$^{-2}$ s$^{-1}$ (integrated over all energies) to $\approx 10^{-5}$ ph cm$^{-2}$ s$^{-1}$; our best-fit Sgr dSph photon flux falls in the middle of this range, $\approx 2\times 10^{-8}$ ph cm$^{-2}$ s$^{-1}$. Then we use our pipeline to recover the flux of the Sgr dSph from the synthetic map, but using a \textit{different} set of templates for Galactic diffuse emission to the ones used to generate the synthetic data. Comparing the recovered Sgr dSph spectrum to the injected one reveals how well our pipeline performs when the input diffuse emission templates are not exactly correct. We carry out this experiment with four diffuse emission template combinations: (1) synthetic data generated from GALPROP + 3D + S, analysed using HD + 3D + S; (2) synthetic data generated from HD + 2D A + S, analysed using HD + 3D + S; (3) synthetic data generated from HD + 3D + S, analysed using HD + 3D + U; (4) synthetic data generated using HD + 3D + S, analysed using HD + 3D but no template for the FBs at all.

We show the results for the first two of these experiments in the two left panels of E.~D.~\autoref{fig:injectionrecovery}; the top left panel shows the recovered energy-integrated photon flux compared to the injected flux, while the bottom left shows the recovered spectra when the input flux is $\approx 2 \times 10^{-8}$ ph cm$^{-2}$ s$^{-1}$. The plot shows that our pipeline yields excellent agreement between the injected and recovered signals for both the integrated flux and the spectrum unless the Sgr dSph signal is $\sim 1$ order of magnitude weaker than our estimate. In no circumstance does our pipeline produce a false signal comparable in magnitude to our observed one. The two right panels of E.~D.~\autoref{fig:injectionrecovery} show the third and fourth tests, where we mismatch the FB template. Here the effects are somewhat larger, but still relatively minor: if we create synthetic data with the S Fermi Bubble template (so that there is structure corresponding to the cocoon), and then analyse it using either the U template or no FB template at all, then we make a factor of $\sim 2-3$ level error in the absolute flux, but no substantial error in the spectral shape. This test suggests that our detection of the Sgr dSph is 
solid, but that we have a factor of $\sim 2-3$ uncertainty in its absolute flux, stemming from our imperfect knowledge of the foreground FBs.

The third validation test we perform is to check whether a fit using the observed stellar distribution of the Sgr dSph as a template performs better than one using a purely geometric template placed at the same position; if the emission really is tracing the stars of the dwarf, and is not merely a chance overlap, a template matching the shape of the dwarf should perform better than a purely geometric distribution. For this purpose we consider disc-shaped templates of varying radii, centred at Galactic coordinates $(\ell,b)=(5.61^\circ, -14.09^\circ)$ --- the dynamical centre of the Sgr dSph -- and repeat our standard procedure of comparing baseline models to baseline + Sgr dSph models, using these geometric templates in place of the Sgr dSph stellar templates. We use our fiducial choices for all other templates (hadronic and bremmstrahlung emission, galactic IC emission, and the Fermi Bubbles).

We show the results of this experiment in S.I.~\autoref{tab:geometric_templates}. We find that geometric templates do perform better than baseline models with no Sgr dSph component, but, as expected, even the best geometric template (for a disc of radius $r=2.0^\circ$) yields significantly less fit improvement ($TS=63.8$) than our fiducial stellar template ($TS=95.2$).
Note that, as we do not nest the geometric and stellar templates in the same analysis, we cannot  translate this difference in test statistic into a formal statistical comparison.
Nevertheless, this difference $\Delta\, TS = 31.4$ is indicative of a preference for the stellar template.
Notice two additional points here:
First, because we tried a wide range of radii for the geometric models, the geometric templates effectively provide an extra degree of freedom that the Sgr dSph template, which is fixed by observations, lacks. Because we fix the template radius while performing each fit, we do not treat the varying radius as an extra degree of freedom when computing the test statistic, but if we did so, then the difference in test statistic between the geometric and stellar templates would be even larger. Second, the geometric model that gives the best fit to the data is in fact the one whose radius most closely approximates the actual size of the core of the Sgr dSph. Indeed, Fig.~\ref{fig:profile_longaxis} (bottom) shows that, in the direction of the short axis, the Sgr dSph stellar profile falls off steeply $\sim 2^\circ-3^\circ$ away from the Sgr dSph centre. Thus the geometric template that gives the best match to the observations happens to be the one that most closely approximates the actual distribution of stars in the Sgr dSph.

\begin{table}[h!]
    \centering
    \small
    \begin{tabular}{llll@{\qquad\qquad}rrrr}
    \hline\hline
    \multicolumn{4}{c}{Template choices} & \multicolumn{4}{c}{Results} \\
    Hadr. / Bremss. & IC & FB & Sgr dSph &
    $-\log(\mathcal{L}_{\rm Base})$  & $-\log(\mathcal{L}_{{\rm Base}+{\rm  Sgr}})$  &  $\mbox{TS}_{\rm Source}$& Significance \\[0.5ex] \hline 
    \multicolumn{8}{c}{Default model} \\[0.5ex]
    HD & 3D & S & Model I & 866680.6 &866633.0  & 95.2  &  $8.1\;\sigma$
    \\[0.5ex]
    HD & 3D & S & Disc ($r=0.5^\circ$) & 866680.6 & 866666.1  & 28.9  &  $3.5\;\sigma$
    \\[0.5ex]
    HD & 3D & S & Disc ($r=1.0^\circ$) & 866680.6 & 866661.3 & 38.6  &  $4.4\;\sigma$
    \\[0.5ex]
    HD & 3D & S & Disc ($r=2.0^\circ$) & 866680.6 & 866648.7 & 63.8  &  $6.3\;\sigma$
    \\[0.5ex]
    HD & 3D & S & Disc ($r=3.0^\circ$) & 866680.6 & 866654.9  &  51.4   &  $5.4\;\sigma$
    \\[0.5ex]
    HD & 3D & S & Disc ($r=4.0^\circ$) & 866680.6 & 866658.1  & 45.0  &  $4.9\;\sigma$
    \\[0.5ex]
    HD & 3D & S & Disc ($r=5.0^\circ$) & 866680.6 &  866661.3 & 38.6 &  $4.4\;\sigma$
    \\[0.5ex]
    HD & 3D & S & Disc ($r=6.0^\circ$) & 866680.6 & 866669.3 & 22.7  &  $2.8\;\sigma$
    \\[0.5ex]
    HD & 3D & S & Disc ($r=7.0^\circ$) & 866680.6 & 866670.4 & 20.4  &  $2.6\;\sigma$
    \\[0.5ex]
    HD & 3D & S & Disc ($r=9.0^\circ$) & 866680.6 & 866664.9 & 31.4  &  $3.7\;\sigma$ \\[0.5ex]
    HD & 3D & S & Disc ($r=11.0^\circ$) & 866680.6 & 866665.8 & 29.6  &  $3.6\;\sigma$ \\[0.5ex]
    HD & 3D & S & Disc ($r=13.0^\circ$) & 866680.6 & 866673.0 & 15.2  &  $1.9\;\sigma$ \\[0.5ex]
    HD & 3D & S & Disc ($r=15.0^\circ$) & 866680.6 & 866676.5 & 8.3  &  $0.9\;\sigma$ \\[0.5ex]
    \hline\hline
    \end{tabular}
    \caption{{\bf Template analysis comparing 
    the result obtained using the fiducial Sgr dSph stellar template 
    to results using disc templates}. The stellar template result 
    is labelled as Model I, top row; 
    subsequent rows are for
    discs templates of various angular radii (as labelled) centred at the dynamical centre of the Sgr dSph. }
    \label{tab:geometric_templates}
\end{table}

Our 
fourth
validation test is to check whether our fit degrades if we artificially rotate or translate the Sgr dSph template; if the signal we are detecting really does come from the Sgr dSph, the best fit should be for a template that traces its actual orientation and position, while rotated or shifted templates should produce progressively worse fits. This check is significant in part because Ref.~\cite{Ackermann2014} performed similar rotation analysis for the hypothesis that the cocoon is tracing a jet from Sgr A$^*$, and found that there was no preference for a jet oriented toward Sgr A$^*$ over one oriented in some other way; they took this as evidence against the jet hypothesis. To check if the Sgr dSph template performs better on this test, we first rerun our analysis pipeline for our default set of templates (first line in \autoref{tab:loglikelihood}), but with the Sgr dSph template rotated about its core. For each rotation angle we compute the TS, and compare to the TS of the original, unrotated model. We plot the result of this experiment in the left panel of E.D.~\autoref{fig:rotationAndTranslationTests}. It is clear that, as expected, the fit is best when we use the actual orientation of the Sgr dSph, and degrades as we increase the rotation. Next, we carry out a similar procedure, but this time rather than rotating the Sgr dSph template about its core, we rotate around the centre of the Galaxy, thereby both translating and rotating the template. (This latter test was motivated by the particular alignment of the Sgr Stream with the previously claimed collimated jets from the Galaxy's supermassive black hole~\cite{Su2012}.) We show the results in the middle panel of E.D.~\autoref{fig:rotationAndTranslationTests}, and, again as expected, the TS strongly favours the true location and orientation of the Sgr dSph. Finally, we translate the Sgr dSph while leaving its orientation unchanged. We show the TS for displaced Sgr dSph in the right panel of E.D.~\autoref{fig:rotationAndTranslationTests}. In this case the fit improves if we do displace the Sgr dSph from its true position by $\approx 4^\circ$ south. The amount by which the shift is favoured is fairly significant -- the TS improved by 40.8, which corresponds to $4.5\sigma$ significance. Interestingly, the direction of the displacement is within a few degrees of the direction anti-parallel to the Sgr dSph's proper motion, suggesting that the dwarf's $\gamma$-ray signal trails it slightly on its orbit.
If IC-emitting CR \pairs \ are largely responsible for the observed \sgr \gr \ signal as suggested by our spectral modelling, a systematic displacement of this signal southward by $\sim 4^\circ$ from the stars of \sgr is quite reasonable as we have explained elsewhere (and see \autoref{sec:CRtrans}).

\section{Transport of IC-emitting CR $e^\pm$}
\label{sec:CRtrans}

We have seen that, while our pipeline detects a signal from the Sgr dSph at very high statistical significance, the fit improves even more (by $\approx 4.5\sigma$) if we displace the Sgr dSph template $\approx 4^\circ$ from its actual position (corresponding to 1.9 kpc at the distance of the Sgr dSph), in a direction very close to anti-parallel to the dwarf's proper motion. Here we demonstrate that a displacement of this type is expected in a model where the $\gamma$-ray signal from the Sgr dSph is powered by MSPs. Part of the MSP signal emerges directly from the MSP magnetospheres, and thus traces the stellar component of the Sgr dSph. However, the majority of the observed signal is, in our model, IC emission powered by \pairs~escaping MSP magnetospheres and interacting with the CMB. The time between when \pairs~leave MSPs and when they IC scatter to produce $\gamma$-ray photons is non-negligible: the CMB is dominated by photons with energies $\sim k_{\rm B}T_{\rm CMB}$ (with $T_{\rm CMB}=2.7$ K), so IC photons with energies of $\sim 1-100$ GeV must be produced by \pairs~with energies $E_{e^\pm} \sim 0.6-6$ TeV. The characteristic IC loss time for such particles is
\begin{equation}
    t_{\rm IC} = \frac{3 m_e^2 c^3}{4 \sigma_{\rm T} E_{e^\pm} U_{\rm CMB}} = 1.2\left(\frac{E_{e^\pm}}{\mbox{TeV}}\right)^{-1}\mbox{ Myr},
\end{equation}
where $m_e$ is the electron mass, $c$ is the speed of light, $\sigma_{\rm T}$ is the Thomson cross section, and $U_{\rm CMB} = a_R T_{\rm CMB}^4 = 0.25$ eV cm$^{-3}$ is the energy density of the CMB.

During this time, the \pairs~will have the opportunity to move a significant distance prior to producing $\gamma$-rays, due to both bulk gas motion and CR flow relative to the gas. 
With regard to bulk advection, we note that the proper speed of the \sgr is $\approx 260$ km s$^{-1}$, and we therefore expect an effective wind of Galactic halo gas to be blowing through (or, at least, around) the dwarf at approximately this speed.
This wind would advect the IC-radiating \pairs \ southward.
Quantitatively, the extent of the angular displacement of an IC \gr \ signal at $E_\gamma$
\begin{equation}
    \Delta \theta_{\rm adv}(E_\gamma) \simeq 1.0^\circ \left(\frac{E_\gamma}{\rm GeV}\right)^{-1} 
    \left(\frac{v_{\rm prop}}{\rm 260 \ km/s}\right)
\end{equation}
where $v_{\rm prop}$ is the proper motion on the sky.
Thus advection is expected to generate a 
southward displacement of $\sim 1^\circ$.

This is less than the displacement we observe, but advection is also likely less important than CR transport through the gas. While the diffusion coefficient for CRs in the galactic halo is very poorly known, we can make an order of magnitude estimate by adopting
the functional form for the diffusion coefficient given in ref~\cite{Gabici2007} which is normalised to $3 \times 10^{27}$ cm$^2$ s$^{-1}$ for a 1 GeV CR in a 3 $\mu$G field. Then the expected diffusive displacement of the IC-radiating \pairs~is
\begin{equation}
    \Delta \theta_{\rm diff}(E_\gamma) \simeq 3.5^\circ \left(\frac{E_\gamma}{\rm GeV}\right)^{-0.12} 
    \left(\frac{B}{\rm 0.1 \ \mu G}\right)^{-0.27} .
\end{equation}
While this is roughly the correct amount of displacement to reproduce what we observe, if the diffusion were isotropic then we would still not have explained the systematic offset between the dwarf and the displaced location picked out by our template analysis. However, we do not expect isotropic diffusion in the environment of the Sgr dSph. Simulations of objects plunging through diffuse halo gas indicate that a generic outcome of such interactions is the development of a coherent magneto-tail back along the objects' direction of motion \cite{Dursi2008}. Such a structure formed by the Sgr dSph plunging through the Milky Way's halo would naturally explain why, rather than being isotropic, the diffusive transport is primarily backwards along the dwarf's trajectory.

\section{Detailed results from Sgr dSph spectral modelling}
\label{ssec:spectralresults}

Detailed results from the spectral modelling described in the Methods are displayed in Supplementary \autoref{fig:FitContours} and
Supplementary \autoref{tab:table1}.

\begin{figure}
    \centering
        \includegraphics[width=0.5\linewidth]{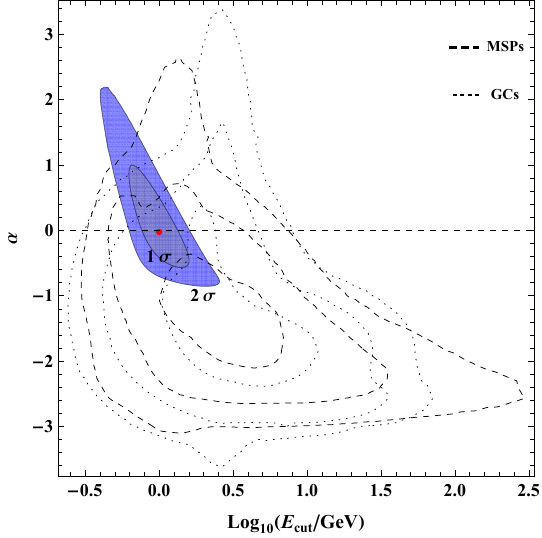}
\caption{ 
Filled contours indicate the best-fit region for the spectral  parameters $E_{\rm cut,prompt}$ and $\alpha$
that determine the shape of the magnetospheric emission from the Sgr dSph; the outer, coloured region shows the 2$\sigma$ region, the inner shows the 1$\sigma$ region, and the red point marks the best fit.
The dotted and dashed contours describe the 1, 2, and 3 $\sigma$ confidence regions measured in ref~\cite{Song2021} for globular clusters (GCs) and individual resolved MSPs, respectively, constructed from the observations as described in Methods.
}
\label{fig:FitContours}
\end{figure}

\begin{table}[ht!]
    \centering
    \begin{tabular}{ccccc}
    \hline
     quantity & best-fit & 68\% c.l.  & units & literature   \\
     &&&& value(s)  \\
    \hline
   $l_0 \equiv L_{\rm \gamma,tot}/M_\star$ & $5.2 $ & $[4.4 , 6.0]$ & $10^{28}$ erg/s/$\msun$ 
    & $\sim (1-10)$ \cite{Sudoh2020}   \\
    $f = L_{\rm \gamma,prompt}/L_{\rm \gamma,IC}$ & $0.83 $ & $[0.59, 1.3]$ & --- &  $\sim 0.1$\cite{Sudoh2020} \\
    $\alpha$ & $0.039 $ & $[-0.38,0.62]$ & --- & $-0.88 \pm 0.44$\cite{Song2021} \\
    $E_{\rm cut, prompt}$ & $1.0 $ & $[0.74, 1.3]$ & GeV & 
    $1.91^{+0.85}_{-0.59} \pm 0.44$ \cite{Song2021} \\
    \hline
    \end{tabular}
    \caption{Best fit spectral parameters with $\pm 1\sigma$ confidence regions as determined from $\chi^2$ fitting to the measured \gr \ spectrum of the Sgr dSph. 
    The parameter $l_0$ is calculated using a stellar mass
    $M_\star = 10^8 \msun$ \cite{Vasiliev2021} for the Sgr dSph.
    See also Supplementary~\autoref{fig:FitContours}
    }
    \label{tab:table1}
\end{table}

\section{Energetics of the Sgr dSph MSP population}
\label{ssec:energetics}

\begin{figure}
 \includegraphics[width=1.\linewidth]{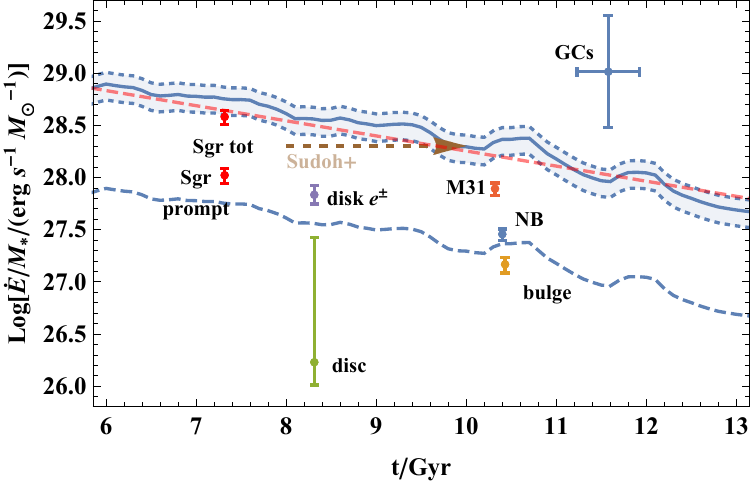}
    \caption{{\bf $\gamma$-ray luminosity per stellar mass  
    versus mean stellar age for a number of stellar systems together with model predictions for the time evolution of the MSP $\gamma$-ray efficiency.}
{\bf Data:}  the data points are for the same systems displayed in main text ~fig.~3.  
The mean stellar ages of these systems have been determined from empirically-determined star formation histories for all these objects (data sources as follows: \sgr \cite{Weisz2014}, M31 \cite{Williams2017}, Galactic Bulge \cite{Bernard2018}, NB \cite{Nogueras-Lara2020} and ref.~\cite{Weisz2013} for the LMC). The globular cluster datum (`GCs') is plotted at the mean measured \gr \ luminosity for the 27 systems analysed in ref.~\cite{Song2021} divided by their stellar masses, and the age is the luminosity-weighted mean age for the 31 systems analysed in ref.~\cite{Wu2022} (while the error bars for this datum show the standard deviations of these measurements for each population). The purple datum shows the secondary electron plus positron luminosity of the Milky Way (`disk $e^\pm$') as inferred in ref.~\cite{Strong2010} and adopting a disk stellar mass of $5.2 \times 10^{10} \ \msun$\cite{Bland_Hawthorn_2016}. {\bf Model curve:} The solid blue curve shows the evolution with time (since the initial, burst-like star formation event) of the total spin-down power generated by a population of MSPs (normalised to the stellar mass expected to host that same population) according to the recent binary population synthesis modelling presented in ref.~\cite{Gautam2021} (with the blue band indicating the estimated the $\pm 1 \sigma$ error on this quantity dominated by the uncertainties in the overall stellar binarity fraction). The {\bf dashed red line} is an approximate fit to the solid blue line described by $5.0 \times 10^{28}$ $ \exp(-t/t_{\rm decay})$ \ erg$/s/\msun$ with $t_{\rm decay} = 3$ Gyr. The dashed blue curve shows 10\% of the mass-normalised spin-down power (with the error band suppressed for clarity). The {\bf brown, dashed, horizontal line} shows the total power (per unit stellar mass) from MSP spin-down we infer from the study by Sudoh et al.~\cite{Sudoh2020} of radio continuum emission from massive, quiescent galaxies (with expected mean stellar ages $>8-10$ Gyr; see the main text for more details).
}
\label{fig:plotLgammaOvrMstarVstSimple}
\end{figure}

As discussed in the main text, 
the \gr \ luminosity per stellar mass we measure for the \sgr \ is substantially brighter than we measure for the Galactic Bulge, Galactic disk, or M31, but is substantially dimmer than is observed for globular clusters.
Indeed, in \autoref{fig:LgammaOvrMstar}, \sgr \ appears as a transition object between gas-poor, low metallicity, low star formation rate, and relatively low stellar mass systems on the left side and relatively gas rich and massive systems (some with appreciable star formation) on the right side.
In order to investigate more deeply how the $\gamma$-ray luminosity of the Sgr dSph compares to that of other observed systems, and to theoretical expectations, in \autoref{fig:plotLgammaOvrMstarVstSimple} we collect measurements of $\gamma$-ray luminosity per unit stellar mass versus approximate age for a range of observed systems, and compare these measurements to model predictions. For the observed systems we include M31, the Milky Way bulge and nuclear bulge, the mean of Milky Way globular clusters, and the Milky Way disc; for the latter we have included both the $\gamma$-ray emission directly measured from MSPs, and the observed \pairs~luminosity of the disc, which may include a significant MSP contribution. As in \autoref{fig:LgammaOvrMstar}, we see that the Sgr dSph is intermediate between the metal-rich galactic systems -- M31, the Milky Way disc and bulge -- and the globular clusters (GCs). However, the figure also reveals a clear trend that galactic systems dim as a function of age, with Sgr dSph as both the youngest and the most luminous of the galactic systems.

The trend with age is consistent with theoretical expectations, indicated by the blue band in Supplementary \autoref{fig:plotLgammaOvrMstarVstSimple} which shows the prediction of a binary population synthesis (BPS) model \cite{Gautam2021} for the total spin-down power per unit stellar mass liberated by magnetic braking of MSPs. Some of this power should emerge as prompt emission, and some as \pairs \ injected into the ISM; the lower dashed blue line shows 10\% of the total spin-down power, a rough estimate for the prompt component. In this particular calculation, the MSPs derive from Accretion Induced Collapse, the population is assumed to be of Solar metallicity, and each binary evolves independently (i.e., the `field star' limit is assumed). Based on the predictions of this model, and the estimated age of the Sgr dSph, we estimate that the $\gamma$-ray signal we have detected can be explained by the presence of $\approx 650$ MSPs in the galaxy. Given that the overall \gr \ luminosity of an MSP population is bounded by the spin-down power, it is evident from the figure that the expected energetics appear to be elegantly sufficient to power the signal from \sgr \ given the (relatively young) mean age of its stars; this age difference naturally explains why the Sgr dSph should be more luminous per unit mass than M31 or components of the Milky Way.

It is also noteworthy that the GCs are considerably more luminous per unit mass than both the BPS model and the Sgr dSph. The extremely high brightness of GCs is plausibly explained by some combination of dynamical effects, which lead to dynamical hardening of binaries and thence a higher production rate of MSPs, and metallicity effects, which lead to higher MSP production because metal-poor stars have weaker winds and thus experience less mass loss during their main sequence lifetimes than Solar-metallicity stars \cite{Ruiter2019}. The former effect would not occur in the Sgr dSph, but the latter would, since the Sgr dSph has a metallicity $\log_{10}(Z_{\rm Sgr}/Z_\odot) \simeq -0.9$ \cite{Vasiliev2020}, where $Z_\odot$ is the solar metallicity, which is comparable to typical GC metallicities.

The final comparison we show in Supplementary \autoref{fig:plotLgammaOvrMstarVstSimple} is with the MSP power inferred by Sudoh et al.\cite{Sudoh2020} in massive, quiescent galaxies ($M_*>10^{9.5}$ M$_\odot$, star formation rate $< 0.1$ M$_\odot$ yr$^{-1}$). Such galaxies typically have stellar population ages $\gtrsim 8-10$ Gyr \cite{McDermid2015,Pacifici2016,Sudoh2020}, and Sudoh et al.~show that they produce anomalously-large synchrotron emission, which they attribute to radiation from \pairs \ injected by MSPs; they infer an injection power $1.8\times 10^{28}$ erg$/$s$/\msun$, which we show as the brown dashed line in Supplementary \autoref{fig:plotLgammaOvrMstarVstSimple}. We see that this estimate is consistent both with the BPS model and comparable to the luminosity we infer for the Sgr dSph.

Our overall conclusion is that the MSP luminosity we have derived for the Sgr dSph is fully consistent with both theoretical expectations and with a wide variety of observed systems. The Sgr dSph is more luminous per unit mass than the Milky Way or M31, but this is easily explained by its youth and low metallicity, and it is comparably- or less-luminous than other observed systems that are of comparable age or metallicity.

\section{Astrophysical \gr \ emission from other dSphs}

On the basis of the normalisation ($L_{\gamma}/M_\star$) supplied by the Sgr dSph $\gamma$-ray detection, we can extrapolate estimates 
for the astrophysical $\gamma$-ray fluxes from a number of other dSph systems, simply by assuming this normalisation applies to them as well; future work should be based on full theoretical models including metallicity and age effects, but the simple calculation we present here can serve as a guide to the system for which such investigations are likely to be fruitful.
Our predictions can, in turn, be compared to i) actual observational upper limits to the \gr \ fluxes from these dSphs  and ii) (model-dependent) predictions for the (WIMP) dark-matter-driven \gr \ fluxes from the same satellite galaxies. For this purpose we use the data assembled in Winter et al.~\cite{Winter2016} for the distances, stellar masses, 
and MSP- and DM-driven fluxes for a population of 30 dSphs satellites of the Milky Way. These authors derive their MSP fluxes by extrapolating the $\gamma$-ray luminosity function of resolved Milky Way MSPs; their result implies that at energies above 500 GeV, galaxies should produce an MSP photon flux per unit stellar mass of $\approx 6.3\times 10^{29}$ s$^{-1}$ M$_\odot^{-1}$, roughly a factor of 40 smaller than the $\approx 2.5\times 10^{31}$ s$^{-1}$ M$_\odot^{-1}$ we detect for the Sgr dSph. We report our revised estimates dwarf spheroidals' MSP luminosity in S.I.~\autoref{tab:dsph_fluxes}. This finding has two implications, which we explore below: first, for some dwarfs this brings the predicted \gr~flux close to current observational upper limits, suggesting that a more detailed analysis of \textit{Fermi}-LAT data might yield a detection. Second, in some dSph galaxies, the predicted MSP flux is comparable to or exceeds the \gr~fluxes that might be expected from dark matter annihilation.

\begin{table}
\begin{tabular}{lcc}
\hline
Galaxy name & Extrapolated MSP flux & Predicted DM flux \\
& (cm$^{-2}$ s$^{-1}$) & (cm$^{-2}$ s$^{-1}$) \\
\hline
\\
Fornax & $2.38\times 10^{-10}$ & $2.06\times 10^{-11}$ \\
Sculptor & $1.10\times 10^{-10}$ & $5.18\times 10^{-11}$ \\
Sextans & $1.96\times 10^{-11}$ & $3.27\times 10^{-11}$ \\
Ursa Minor & $1.95\times 10^{-11}$ & $8.20\times 10^{-11}$ \\
Leo I & $1.59\times 10^{-11}$ & $6.52\times 10^{-12}$ \\
Draco & $1.18\times 10^{-11}$ & $8.20\times 10^{-11}$ \\
Carina & $8.12\times 10^{-12}$ & $1.64\times 10^{-11}$ \\
Leo II & $4.54\times 10^{-12}$ & $5.18\times 10^{-12}$ \\
Bootes I & $1.36\times 10^{-12}$ & $2.06\times 10^{-11}$ \\
Canes Ven. I & $1.33\times 10^{-12}$ & $6.52\times 10^{-12}$ \\
Ursa Major II & $1.10\times 10^{-12}$ & $2.59\times 10^{-10}$ \\
Reticulum II & $5.27\times 10^{-13}$ & $2.59\times 10^{-10}$ \\
Coma Ber. & $5.19\times 10^{-13}$ & $1.30\times 10^{-10}$ \\
Hercules & $4.48\times 10^{-13}$ & $1.64\times 10^{-11}$ \\
Ursa Major I & $4.25\times 10^{-13}$ & $2.59\times 10^{-11}$ \\
Tucana III & $2.67\times 10^{-13}$ & $2.59\times 10^{-10}$ \\
Grus II & $2.53\times 10^{-13}$ & $6.52\times 10^{-11}$ \\
Tucana IV & $1.99\times 10^{-13}$ & $6.52\times 10^{-11}$ \\
Tucana II & $1.88\times 10^{-13}$ & $8.20\times 10^{-11}$ \\
Eridanus II & $1.60\times 10^{-13}$ & $2.59\times 10^{-12}$ \\
Willman I & $1.45\times 10^{-13}$ & $1.64\times 10^{-10}$ \\
Segue 1 & $1.34\times 10^{-13}$ & $4.11\times 10^{-10}$ \\
Leo IV & $7.53\times 10^{-14}$ & $1.03\times 10^{-11}$ \\
Horologium I & $6.65\times 10^{-14}$ & $3.27\times 10^{-11}$ \\
Phoenix II & $6.55\times 10^{-14}$ & $3.27\times 10^{-11}$ \\
Canes Ven. II & $6.51\times 10^{-14}$ & $1.03\times 10^{-11}$ \\
Reticulum III & $4.95\times 10^{-14}$ & $2.06\times 10^{-11}$ \\
Columba I & $3.91\times 10^{-14}$ & $5.18\times 10^{-12}$ \\
Indus I & $3.50\times 10^{-14}$ & $2.59\times 10^{-11}$ \\
Indus II & $2.24\times 10^{-14}$ & $3.27\times 10^{-12}$ \\
\hline
\end{tabular}
\caption{
\label{tab:dsph_fluxes}
Extrapolated MSP photon and predicted dark matter annihilation photon fluxes at energies $E_\gamma > 500$ MeV from nearby dSph galaxies, taken from the sample of ref.~\cite{Winter2016}. Column 1: galaxy name; column 2: predicted MSP photon flux based on the Sgr dSph (see SI for details); column 3: DM annihalation flux predicted by ref.~\cite{Winter2016}.
}
\end{table}

\subsection{Comparison with existing upper bounds}

To estimate whether other dwarf spheroidals might be detectable, we compare our differential flux predictions (incorporating both prompt and IC emission where, for simplicity, we make the approximation that the CMB-dominated ISRF of the \sgr \ also pertains in each other dSph under consideration)
against the results from ref.~\cite{Mazziotta2012}\footnote{This is the most recent publication we can find that explicitly tabulates bin-by-bin, numerical 95\% confidence upper limits on the differential flux received by a number of dSphs that also appear in the compilation of ref.~\cite{Winter2016}}. On the basis of this comparison, we do not predict \gr \ emission from any dSph that surpasses the upper limits from ref.~\cite{Mazziotta2012}. However two dSphs reach a significant fraction of the relevant upper limit in at least one energy bin (of width $log_{10}(\Delta E$/GeV) = 0.5): Fornax (which reaches 0.24 of the upper limit for the energy bin centred at 1.36 GeV) and Sculptor (which reaches 0.09 of the upper limit for the energy bin centred at 2.46 GeV). Furthermore, the results of ref.~\cite{Mazziotta2012} were obtained using Pass7 \fermi \ data accumulated over only the first 3 years of \fermi \ operation. On the basis of, e.g., the results of ref.~\cite{Ackermann2015} we expect that updated upper limits (Pass8, 15 years data) should be at least a factor of 4 more stringent. This makes Fornax and Sculptor both very interesting targets for a future study, though we remind the reader that our predictions are predicated on a normalisation obtained from the \sgr \ detection that may be somewhat over-optimistic because it ignores the stellar age effects evidenced in Supplementary \autoref{fig:plotLgammaOvrMstarVstSimple}\footnote{The stellar population of Sculptor, in particular, is significantly older \cite{Bettinelli2019} than that of Sagittarius, giving the MSP population more time to have spun down, though Fornax, on the other, has experienced some significant and relatively recent star formation \cite{Rusakov2021}, like Sagittarius, qualifying it as a particularly compelling target for \gr \ observation.}. After these two, the brightest expected  dSphs are Sextans, Ursa Minor, Leo I, and Draco. These may also be interesting targets, though we note that we expect that they are almost one order of magnitude dimmer than Fornax and Sculptor.

\subsection{Comparison with predicted DM annihilation fluxes}

Winter et al.\cite{Winter2016} estimate DM annihilation fluxes for nearby dSphs using a DM annihilation cross section derived by assuming that the the Galactic Centre Excess (GCE) is a DM signal. We caution that this is likely only an upper limit, since of course our finding for the Sgr dSph suggests that some or all of the GCE is in fact due to MSPs (see also ref.~\cite{Gautam2021}). Nonetheless, we proceed with our calculation using the Winter et al.~estimate precisely because it represents an upper limit on the DM signal. Comparing the MSP and DM signals estimated in S.I.~\autoref{tab:dsph_fluxes} leads us to the important finding that, in contrast to the results obtained by Winter et al., there are three dSphs for which the MSP-driven $> 500$ MeV photon number flux exceeds the predicted DM flux (viz., Fornax by $\sim$12; Leo I by $\sim$2.4; and Sculptor by $\sim$2.1) and three more where it exceeds $\sim 1/2$ the DM flux (viz., Leo \
II with 0.89; Sextans with 0.60; and Carina with 0.50 of the DM flux).
A clear implication of these, albeit preliminary, results is that these targets should be avoided in the quest to better constrain putative WIMP DM self-annihilation cross-sections. By contrast, there remain other dSphs where the expected DM signal remains comfortably much larger than the MSP signal; these are more promising targets.

\clearpage

\printbibliography[segment=\therefsegment,title={Supplementary Information References}, check=onlynew]

\end{document}